\documentclass[main]{aa}  

\usepackage{graphicx}
\usepackage{txfonts}
\usepackage[colorlinks=true,citecolor=blue,linkcolor=blue,urlcolor=blue]{hyperref}

\newcommand{\Msun}{\,$M_{\odot}$}       
       
\newcommand{\um}{\,$\mu$m}      
    
\newcommand{\qpah}{q$_{\rm PAH}$}       
\newcommand{\logoh}{$12+\log{\rm{(O/H)}}$}      
\newcommand{\fobsc}{$f_{\rm obsc}$}
\newcommand{\Ssfr}{$\Sigma_{\rm SFR}$}

\begin{document} 

   \title{A new census of dust and polycyclic aromatic hydrocarbons at $z=0.7-2$ with JWST MIRI
   }
   \titlerunning{Dust and PAHs at $z=0.7-2.0$}

   \author{Irene Shivaei
          \inst{1,2}
          \and
          Stacey Alberts\inst{2} \and
          Michael Florian\inst{2} \and
          George Rieke\inst{2} \and
          Stijn Wuyts\inst{3} \and
          Sarah Bodansky\inst{4}\and
          Andrew J. Bunker\inst{5}\and
          Alex J. Cameron\inst{5}\and
          Mirko Curti\inst{6}\and
          Francesco D'Eugenio\inst{7,8} \and
          Ugnė Dudzevičiūtė\inst{9} \and
          Zhiyuan Ji\inst{2} \and
          Benjamin D. Johnson\inst{10} \and
          Ivan Kramarenko\inst{11} \and
          Jianwei Lyu\inst{2} \and
          Jorryt Matthee\inst{10} \and
          Jane Morrison\inst{2} \and
          Rohan Naidu\inst{12}\and
          Pablo G. P\'erez-Gonz\'alez\inst{1} \and
          Naveen Reddy\inst{13} \and
          Brant Robertson\inst{14} \and
          Yang Sun\inst{2} \and
          Sandro Tacchella\inst{7,8} \and
          Katherine Whitaker\inst{4,15}\and
          Christina C. Williams\inst{16} \and
          Christopher N. A. Willmer\inst{2} \and
          Joris Witstok\inst{7,8} \and
          Mengyuan Xiao\inst{17}\and
          Yongda Zhu\inst{2,13}
          }

   \institute{{Centro de Astrobiolog\'{i}a (CAB), CSIC-INTA, Carretera de Ajalvir km 4, Torrej\'{o}n de Ardoz, 28850, Madrid, Spain}\\
              \email{ishivaei@cab.inta-csic.es}
         \and
             {Steward Observatory, University of Arizona, Tucson, AZ 85721, USA}
        \and
            {Department of Physics, University of Bath, Claverton Down, Bath, BA2 7AY, UK}
        \and    
            {Department of Astronomy, University of Massachusetts, Amherst, MA 01003, USA}
        \and
            {Department of Physics, University of Oxford, Denys Wilkinson Building, Keble Road, Oxford OX1 3RH, UK}
        \and
            {European Southern Observatory, Karl-Schwarzschild-Strasse 2, 85748 Garching, Germany }
        \and
            {Kavli Institute for Cosmology, University of Cambridge, Madingley Road, Cambridge, CB3 0HA, UK}
        \and
            {Cavendish Laboratory, University of Cambridge, 19 JJ Thomson Avenue, Cambridge, CB3 0HE, UK}
        \and
            {Max-Planck-Institut f\"{u}r Astronomie, K\"{o}nigstuhl 17, 69117 Heidelberg, Germany}
        \and    
            {Center for Astrophysics $|$ Harvard \& Smithsonian, 60 Garden St., Cambridge MA 02138 USA}
        \and
            {Institute of Science and Technology Austria (ISTA), Am Campus 1, 3400 Klosterneuburg, Austria}
        \and
            {MIT Kavli Institute for Astrophysics and Space Research, 77 Massachusetts
Avenue, Cambridge, MA 02139, USA}
        \and
            {Department of Physics and Astronomy, University of California, Riverside,
900 University Avenue, Riverside, CA 92521, USA}
        \and
            {Department of Astronomy and Astrophysics, University of California, Santa Cruz, 1156 High Street, Santa Cruz, CA 95064, USA}
        \and
            {Cosmic Dawn Center (DAWN), Niels Bohr Institute, University of Copenhagen, Jagtvej 128, København N, DK-2200, Denmark}
        \and
            {NSF’s National Optical-Infrared Astronomy Research Laboratory, 950 North Cherry Avenue, Tucson, AZ 85719, USA}
        \and
            {Department of Astronomy, University of Geneva, Chemin Pegasi 51, CH1290 Versoix, Switzerland}
        }

   \date{Received 02/2024; accepted 10/06/2024}

 \abstract
 {}
 {This paper utilises the \textit{James Webb} Space Telescope (JWST) Mid-Infrared Instrument (MIRI) to extend the observational studies of dust and polycyclic aromatic hydrocarbon (PAH) emission to a new mass and star formation rate (SFR) parameter space beyond our local Universe. The combination of fully sampled spectral energy distributions (SEDs) with multiple mid-infrared (mid-IR) bands and the unprecedented sensitivity of MIRI allows us to investigate dust obscuration and PAH behaviour from $z=0.7$ up to $z=2$ in typical main-sequence galaxies. Our focus is on constraining the evolution of PAH strength and the dust-obscured luminosity fraction before and during cosmic noon, the epoch of peak star formation activity in the Universe.}
 {In this study, we utilise MIRI multi-band imaging data from the SMILES survey (5 to 25\,{\um}), complemented with NIRCam photometry from the JADES survey (1 to 5\,{\um}), available HST photometry (0.4 to 0.9\,{\um}), and spectroscopic redshifts from the FRESCO and JADES surveys in GOODS-S for 443 star-forming (without dominant active galactic nucleus (AGN)) galaxies at $z=0.7-2.0$. This redshift range was chosen to ensure that the MIRI data cover mid-IR dust emission. 
 Our methodology involved employing ultraviolet (UV) to IR energy balance SED fitting to robustly constrain the fraction of dust mass in PAHs and dust-obscured luminosity. 
 Additionally, we inferred dust sizes from MIRI 15\,{\um} imaging data, enhancing our understanding of the physical characteristics of dust within these galaxies.}
 {We find a strong correlation between the fraction of dust in PAHs (PAH fraction, {\qpah}) with stellar mass. Moreover, {the sub-sample with robust {\qpah} measurements ($N=216$) shows a similar behaviour between {\qpah} and gas-phase metallicity to that at $z\sim 0$}, suggesting a universal relation: {\qpah} is constant ($\sim 3.4$\%) above a metallicity of $Z \sim 0.5\,Z_{\odot}$ and decreases to $<1$\% at metallicities $\lesssim 0.3\,Z_{\odot}$. This indicates that metallicity is a good indicator of the interstellar medium properties that affect the balance between the formation and destruction of PAHs. The lack of a redshift evolution from $z\sim 0-2$ also implies that above $Z\sim 0.5\,Z_{\odot}$ the PAH emission effectively traces obscured luminosity and the previous locally calibrated PAH-SFR calibrations remain applicable in this metallicity regime.
 We observe a strong correlation between the obscured UV luminosity fraction (ratio of obscured to total luminosity) and stellar mass. Above the stellar mass of $M_*>5\times 10^9$\,{\Msun}, on average, more than half of the emitted luminosity is obscured, while there exists a non-negligible population of lower-mass galaxies with $>50\%$ obscured fractions. At a fixed mass, the obscured fraction correlates with SFR surface density. This is a result of higher dust covering fractions in galaxies with more compact star-forming regions. Similarly, galaxies with high IRX (IR to UV luminosity) at a given mass or UV continuum slope ($\beta$) tend to have higher {\Ssfr} and shallower attenuation curves, owing to their higher effective dust optical depths and more compact star-forming regions.}
 {}

   \keywords{
               }

   \maketitle

\section{Introduction} \label{sec:intro}

Dust stands out as one of the most intriguing baryonic components in galaxies. While it constitutes only $\sim 0.001-0.01\%$ of the baryonic matter by mass, its emission contributes to about half of the electromagnetic content of our Universe that comes from galaxy formation and evolution processes \citep{dole06}. Dust is essential to the chemistry and physics of the interstellar medium (ISM), and, as has traditionally been recognised, its attenuation effects are significant in the ultraviolet (UV) and optical wavelengths \citep[][among many more]{cardelli89,gordon97,weingartner01a,calzetti00,salim20}. 

Among different components of interstellar dust, polycyclic aromatic hydrocarbons (PAHs) stand out, as not only are they abundant throughout the Universe, they influence the chemistry and thermal budget of the ISM by controlling the ionisation balance, dominating the heating process in neutral gas through photoelectric heating \citep{bakes94,wolfire95,helou01}, and acting as catalysts for the formation of H$_2$ molecules \citep{thrower12,boschman15,Barrera23}. The PAHs undergo transient heating by absorbing single UV photons and emit light in the mid-infrared (mid-IR) range from $\sim 3-20${\um}, with the strongest emission at 7.7\,{\um}. Their broad emission features are widely observed in the spectra of star-forming galaxies, contributing to up to 20\% of the total IR emission \citep{lagache04,smith07,tielens08,li20}. 

Polycyclic aromatic hydrocarbons have been studied extensively in the local Universe. Their mass or luminosity fraction relative to that of the total dust is observed to decrease below a certain gas-phase metallicity of {\logoh}$\sim 8.1-8.3$, depending on different calibrations \citep{engelbracht05,madden06,draine07b,marble10,remyruyer15,aniano20}. 
The underlying physical cause of this behaviour is generally attributed to either a lack of production or preferential destruction at low metallicities. The reduced PAH abundance at low metallicities may be due to a lower abundance of carbon in the gas phase preventing PAH formation in the ISM \citep{draine07b} or a deficiency of stars that produce PAHs (such as AGB stars; \citealt{galliano08}). It may also be due to a rapid destruction of PAHs by the more intense and harder UV radiation in the low-metallicity ISM systems in which dust shielding is reduced \citep{madden06,hunt10,xie19} or by thermal sputtering in shock-heated gas with a lower cooling rate because of low metallicity \citep{li20}. The increased abundance of PAHs at higher metallicities is also attributed to an accelerated production path through shattering, which becomes more efficient above a certain metallicity \citep{seok14}.

Owing to its prevalence and being accessible by various IR telescopes, the $\sim 7-8$\,{\um} mid-IR emission is calibrated as a practical and useful indicator of obscured luminosity and the star formation rate (SFR) in the metal-rich star-forming galaxies regime \citep{peeters04,calzetti07,kennicutt09,rujopakarn13,shipley16}.
This calibration is often combined with unobscured tracers, such as UV and optical emission lines, to provide a total SFR estimation \citep{kennicutt12}. At $z\sim 1-3$, known as cosmic noon, the 7.7\,{\um} PAH emission was captured by \textit{Spitzer}/MIPS 24\,{\um} \citep{rieke04}, making it a widely used tracer of obscured luminosity and SFR to advance our knowledge about SFR evolution and the star-forming main sequence \citep{reddy08,elbaz11,rujopakarn13,whitaker12b,shivaei15a,shivaei17}, obscured fraction \citep{reddy10,whitaker17}, and IR luminosity function \citep{perez-gonzalez05,lefloch05,magnelli11}, as well as to study PAH characteristics themselves \citep{shivaei17}. 
However, MIPS confusion noise \citep{dole04} inevitably constrained these studies to more IR-bright, dusty, and massive galaxies at cosmic noon. While some studies employ sophisticated stacking techniques to extend the dynamic range to lower-mass galaxies, this approach by definition sacrifices information about the properties of individual galaxies.

The perspective on mid-IR dust emission in galaxies beyond the local Universe has undergone a significant shift with the advent of the Mid-Infrared Instrument (MIRI; \citealt{rieke15,wright23}) on the \textit{James Webb} Space Telescope (JWST; \citealt{gardner23}). This instrument offers an order-of-magnitude improvement in spatial resolution without the confusion noise limitation at comparable wavelengths to MIPS. 
Moreover, its continuous wavelength coverage from 5 to 25{\um} with nine intermediate-width photometric bands provides a leap forward in characterising sources compared to its predecessors. This allows for novel studies of dust, both in the local Universe \citep{alvarez23,sandstrom23,chastenet23,armus23} and beyond \citep{kirkpatrick23,lyu24,lin24,perez-gonzalez24}.
Importantly, the high angular resolution enables not only the detection of dust emission from individual typical galaxies at $z>0.5$, but also, in many cases, the measurement of the spatial extent and morphology of the dust within the galaxies \citep{shen23,magnelli23} -- providing new opportunities to explore the dust content of galaxies all the way to cosmic noon.

Here, we present the first paper of the {\sc{Systematic Mid-infrared Instrument Legacy Extragalactic Survey}} (SMILES; GTO 1207) on the PAH and dust emission of star-forming galaxies at $z=0.7-2.0$. SMILES is the widest MIRI survey in Cycles 1 and 2 of JWST observations and covers the full MIRI photometric range from 5.6 to 25.5{\um}. The comprehensive multi-wavelength coverage facilitates a robust investigation of PAHs and dust emission in one of the richest extragalactic deep fields, the Great Observatories Origins Deep Survey South (GOODS-S; \citealt{dickinsongoods03}). The structure of the paper is as follows. In Section~\ref{sec:method}, we detail the dataset used and the methodologies employed to estimate various physical properties of galaxies, including the spectral energy distribution (SED) fitting. Section~\ref{sec:pahs} presents the results on the evolution of PAH mass and luminosity fraction from $z\sim 0$ to 2. In Section~\ref{sec:obsc}, we investigate the obscured UV luminosity fraction of galaxies as a function of mass and SFR surface density. The findings are summarised in Section~\ref{sec:summary}. Throughout this paper, we assume a $\Lambda$CDM flat cosmology with $H_0=70$\,km\,s$^{-1}$\,Mpc$^{-1}$ and $\Omega_{\Lambda}=0.7$, and a \cite{chabrier03} initial mass function (IMF).

\begin{figure*}
\centering
        \includegraphics[width=\textwidth]{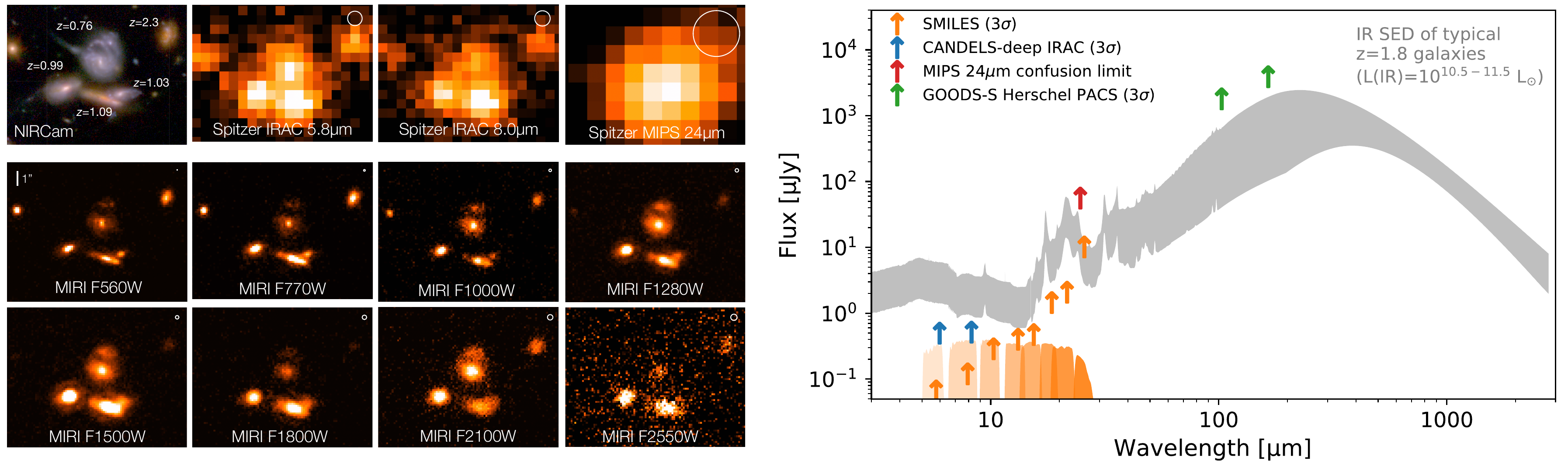}
    \caption{
    Power of SMILES in studying the mid-IR dust properties of galaxies at cosmic noon. 
    Left: The top-left image shows a group of galaxies at $z\sim 0.7-2.3$ (matching the redshift range in this paper) in F444W/F200W/F090W NIRCam colours (JADES). The pre-JWST mid-IR view of these galaxies is shown in the top row from \textit{Spitzer} IRAC and MIPS (GOODS survey; \citealt{dickinson07,ashby15}). The bottom two rows show the mid-IR images with JWST MIRI at 5.6, 7.7, 10, 12.8, 15, 18, 21, and 25.5\,{\um} (SMILES). The PSF sizes are shown with white circles in the corners (diameter of the circle is the FWHM).
    The unprecedented improvement of MIRI in spatial resolution (0.19 to 0.86$''$ from 5.6 to 25.5\,{\um}) compared to that of IRAC (1.72 and 1.88$''$ at 5.8 and 8\,{\um}) and MIPS 24\,{\um} (6$''$) makes it possible to observe the mid-IR emission of individual typical galaxies at these redshifts.
    Right: The IR dust SEDs of typical galaxies at $z=1.8$ with IR luminosities of $10^{10.5}-10^{11.5}\,L_{\odot}$ (\citealt{rieke09}; corresponding to obscured SFRs of $\sim 5-50$\,{\Msun}/yr assuming the \citealt{kennicutt12} calibration) in grey shaded region, along with 3$\sigma$ sensitivity limits of the GOODS-S \textit{Herschel}/PACS \citep[green;][]{magnelli13}, \textit{Spitzer}/MIPS 24\,{\um} confusion limit (red; 56\,$\mu$Jy, \citealt{dole04}), \textit{Spitzer} IRAC Channels 3 and 4 \citep[blue;][]{guo13}, and the MIRI sensitivities of SMILES in $\sim 10-30$\,min exposures (orange; Section~\ref{sec:smiles}). MIRI filter transmission curves are shown with orange shaded regions. The high sensitivity and multi-band capability of MIRI enables measurements and characterisations of IR emission of typical galaxies at cosmic noon.
    }
    \label{fig:SMILES}
\end{figure*}

\section{Methods and observations} \label{sec:method}

\subsection{SMILES survey (MIRI data)} \label{sec:smiles}
This paper is based on the MIRI imaging survey of {\sc{Systematic Mid-infrared Instrument Legacy Extragalactic Survey}} (SMILES; \citealt{rieke24,alberts24}). SMILES is a MIRI US-GTO programme (PID 1207, PI: George~Rieke) that surveys 34 arcmin$^2$ (using 15 MIRI pointings in a 3$\times$ 5 mosaic) centred on the Hubble Ultra Deep Field (HUDF; \citealt{beckwith06}). The survey is centred at 03h32m34.80s -27d48m32.60s and includes the eight broad bands of the MIRI Imager from 5.6 to 25.5{\um}. A standard MIRI 4-point dither pattern was adopted to improve the point spread function (PSF) sampling and mitigate detector artifacts and cosmic-ray events. We used a FASTR1 readout pattern. Each pointing had a total science exposure time of 2.17 hours, divided over eight bands. The exposure times per band were designed to optimise the detection of typical active galactic nucleus (AGN) and star-forming galaxies down to an SFR of $\sim 10$\Msun yr$^{-1}$.
The observations were carried out in December 2022 and January 2023, which achieved a 5$\sigma$ point source sensitivity of 0.11\,$\mu$Jy (AB mag $=$ 26.3), 0.20\,$\mu$Jy (25.6), 0.48 (24.7), 0.67 (24.3), 0.79 (24.2), 2.42 (22.9), 3.47 (22.6), and 16.91 (20.8) in F560W (754\,s exposure time), F770W (866\,s), F1000W (644\,s), F1280W (755\,s), F1500W (1121\,s), F1800W (755\,s), F2100W (2186\,s), and F2550W (832\,s)\footnote{Our derived sensitivities are $\sim 2-3\times$ deeper than the sensitivities predicted by ETC (version 3.0) for our observation setup.}. 
The achieved sensitivities are an order of magnitude deeper than the confusion limit of \textit{Spitzer}/MIPS 24\,{\um} (56\,$\mu$Jy; \citealt{dole04}) at similar wavelengths. The optimal combination of depth, spatial resolution, coverage area, and multi-wavelength capabilities makes SMILES well suited to investigating mid-IR dust emissions in galaxies during the cosmic noon epoch. In Figure~\ref{fig:SMILES}, we show the advancement of this dataset compared to the existing IR data to study cosmic noon galaxies.

\paragraph{Data reduction} Data reduction of MIRI images was performed using the nominal JWST Pipeline released by the Space Telescope Science Institute (JWST Calibration Pipeline
v1.10.0; \citealt{bushouse_v1100}), using JWST Calibration Reference System (CRDS) {\tt jwst 1084.pmap}, with two custom modifications on the background subtraction and astrometry, and a post-processing flux calibration correction on F560W and F770W (see below). After the astrometric and background corrections, the final MIRI mosaic was made through the Pipeline stage 3 with a pixel scale of 0.06''.

\paragraph{Background correction} MIRI photometry relies on background-limited sensitivity across all bands, but particularly at longer wavelengths at which the thermal emission from the telescope is significant \citep{glasse15,rigby23}. The background imprints a large spatial gradient on the image \citep{dicken24}. In addition, in some cases there are tree-ring-shaped features and striping along the detector columns and rows \citep{morrison23} that can also be corrected for during the background subtraction step. Therefore, proper correction is crucial to reach maximum sensitivity. We adopted a custom background subtraction methodology that is explained in detail in \citet{perez-gonzalez24} and \citet{alberts24}. In brief, this routine was performed on the products of the {\sc calwebb image2} (cal files) step of the pipeline. It combines the cal images taken in the same band from different pointings to create a `super-background', departing from the routine in the JWST pipeline, which utilises only the four dithers for each pointing. Through an iterative process, for each cal file, we masked out sources, large gradients, and striping using median filtering. This way, the super-background was constructed for each uncleaned cal file. Finally, a median subtraction was applied to remove any remaining background variation, usually caused by cosmic ray showers.

\paragraph{Astrometry correction}
This step was done on the background-subtracted cal files, the output from the previous paragraph. The {\tt tweakreg} step in the Pipeline (in stage 3) was designed to calculate the astrometry correction of the image. However, in the earlier versions it was designed for automated processing without a user input reference catalogue. As the MIRI  images  do not cover an area of the sky  large enough to contain a sufficient number of \textit{Gaia} point sources, it is necessary to supply an external reference catalogue made from the \textit{Hubble} Space Telescope (HST), or NIRCam images in the field were necessary. We adopted the {\tt tweakwcs} package with a custom routine outside of the Pipeline.
We used the NIRCam F444W mosaic (JADES survey; registered to \textit{Gaia} DR3) to correct the F560W image by creating matched catalogues of high signal-to-noise (S/N) F444W and F560W sources. We used the registered F560W to correct F770W, and so on, up to F1500W. The F1500W was used to correct F1800W, F2100W, and F2550W images as the number of high S/N sources in each band was insufficient for independent corrections. 
The achieved astrometric accuracy is $0.01-0.02$'' (1$\sigma$) in all filters except for F2550W, which has a low number of high S/N sources and an astrometric accuracy of $\sim 0.04$''.

\paragraph{Flux calibration correction}
After data reduction and photometry (see below), we realised the F560W and F770W flux measurements are systematically overestimated. This became apparent when the predicted F560W fluxes from full SED fitting to HST, NIRCam, and MIRI data (excluding F560W) of sources with spectroscopic redshifts were consistently lower than the observed F560W photometry. Moreover, we compared the photometry with \textit{Spitzer} IRAC and found a similar systematic overestimation in MIRI photometry.
Based on simulations of the PSF, we concluded that this is likely due to an underestimation of the cruciform spikes of the MIRI PSF \citep{gaspar21} in the flux calibration step of the Pipeline (this is now updated in the Pipeline; see below). Therefore, we calculated correction factors by comparing the observed F560W and F770W fluxes with the predicted fluxes from best-fit SEDs of 25 isolated stars using ASTRODEEP photometry up to IRAC2 \citep{lyu24}.
The correction factors of 1.26 and 1.04 were applied to the F560W and F770W Kron fluxes, respectively (see below for photometry). 
The absolute flux calibration was updated in the JWST Calibration Reference Data System (CRDS) after the current reductions on December 14 2023, to better account for the cruciform effect and the time dependence of MIRI sensitivity (Gordon et~al., in prep). The correction factors relevant to SMILES are 1.27 and 1.18 in F560W and F770W, respectively. While the F560W correction is in agreement with our calculations, the official new calibrations for F770W are higher than ours. However, since the presented work is not highly dependent on the F770W flux (rest-frame of $\sim 2.5-4.5$\,{\um}), our results will be unchanged with either correction factor.

\paragraph{Photometry} \label{sec:miri-phot}
MIRI source detection and photometric extraction were performed using a modified version of the JADES photometric pipeline ({\tt jades-pipeline}, for the pipeline see \citealt{rieke23,robertson23}; for MIRI photometry see also \citealt{lyu24}). We briefly summarise the main points here: a detection and S/N map were constructed by stacking the F560W and F770W mosaics, which were then used to define an initial blended segmentation map down to a low threshold in source S/N.  This segmentation map was then processed to optimally deblend sources and remove spurious noise spikes.  The final segmentation map and detection image were then used to define source centroids and photometry was measured in all filters at these centroids.  In this work, we adopted 2.5$\times$ scaled Kron photometry.  As was noted above, correction factors were applied to the flux calibration at F560W and F770W. Aperture corrections were determined as the fraction of flux outside a given Kron aperture based on empirical PSFs for F560W and F770W (using comissioning data PID 1028, \citealt{dicken24}, A. G{\'a}spar, private communication) and model PSFs generated from WebbPSF \citep{perrin14} for the other bands.  Photometric uncertainties were derived by placing apertures at random positions on a masked mosaic, which accounts for correlated pixel noise. A more detailed description of the construction of the SMILES photometric catalogue is presented in \cite{alberts24}.

\begin{figure*}
        \centering
        \includegraphics[width=.55\textwidth]{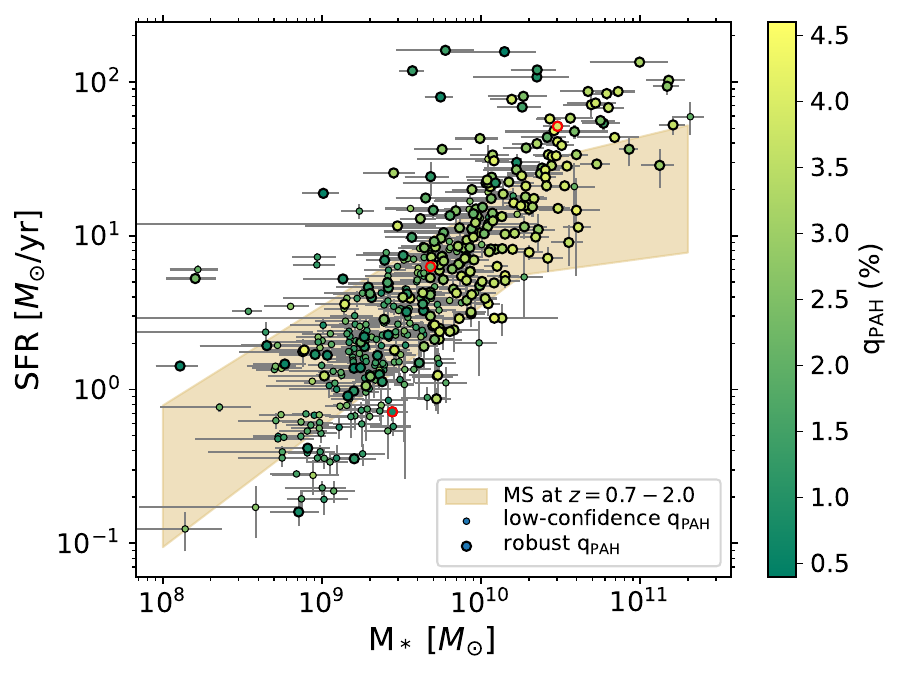} \quad
        \includegraphics[width=.4\textwidth]{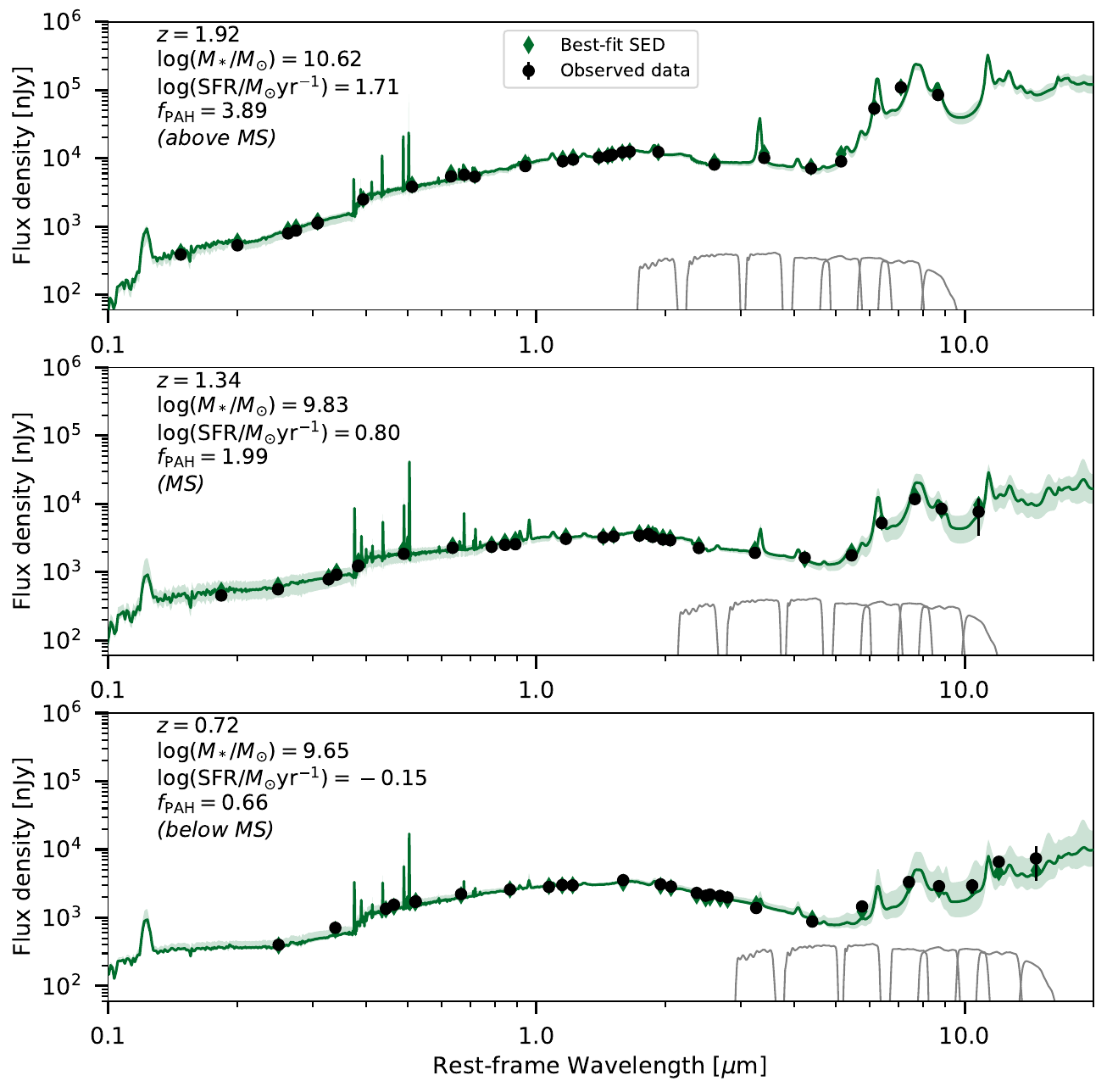}
    \caption{
Left panel: SFR versus stellar mass of the sample. The sample is divided into those with robust and low-confidence PAH mass fractions (same as in Figure~\ref{fig:sample}). The shaded region indicates star-forming main sequence at the redshift range of our galaxies, $z=0.7-2.0$, derived from the PROSPECTOR-inferred SFR and mass of \cite{leja22}, same as the SFR and stellar masses inferred in this work. Three SED examples of the symbols with red edges are shown on the right.
Right panel: Examples of the SEDs of three galaxies that are above, on, and below the main sequence, shown with the red edges in the left plot. The MIRI filter transmission curves are also shown. The wealth of data from HST, NIRCam, and MIRI provides robust SED fits for this sample at $z=0.7$ to 2.0. 
More examples of the SED fits are provided in the online repository: \href{https://zenodo.org/records/12671075}{https://zenodo.org/records/12671075} .
}
    \label{fig:MS-SEDs}
\end{figure*}
\subsubsection{JADES, JEMS, and HST data}

We used deep NIRCam imaging data in multiple filters spanning 1-5\,{\um} from the JWST Advanced Deep Extragalactic Survey (JADES) in GOODS-S (PID 1180 , PI: D. Eisenstein) and the medium bands of the JWST Extragalactic Medium-band Survey (JEMS, PID 1963, \citealt{williams23}). We refer to the JADES survey data release papers for further details on the observations and data reduction \citep{rieke23,eisenstein23}. In brief, JADES GOODS-S covers a total of 67 square-arcminutes. The survey has two tiers: i) Deep with 27 square-arcminutes in NIRCam F090W, F115W, F150W, F200W, F277W, F335M, F356W, F410M, and F444W bands, and ii) Medium with 40 square-arcminutes with the same filters, but without F335M. We also included NIRCam medium bands F210M, F430M, F460M, and F480M from the JEMS survey, where available.

We used Kron photometry from the custom-developed {\tt jades-pipeline} \citep[discussed in Section~\ref{sec:miri-phot};][]{robertson23,rieke23}. 
Furthermore, we used the spectroscopic redshifts from JADES NIRSpec \citep{bunker23}, where available (Section~\ref{sec:sed-fitting-other}).
To cover the shorter wavelengths, we also included existing HST/ACS F435W, F606W, F775W, F814W, and F850LP data from the Hubble Legacy Fields (HLF) v2.0 \citep{illingworth13}, applying the same photometry {\tt jades-pipeline} as for the NIRCam data.

\begin{figure*}
\centering
        \includegraphics[width=.9\textwidth]{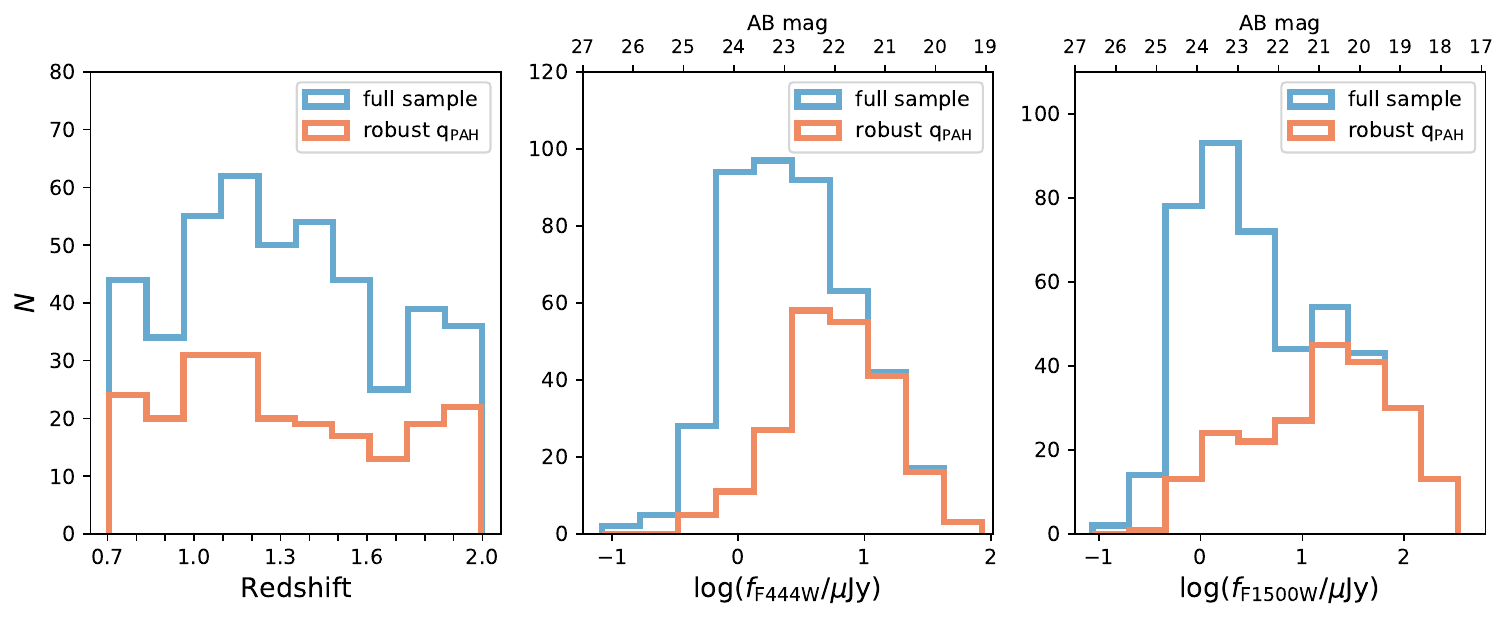}
    \caption{Distributions of redshift (left), F444W flux density (middle), and F1500W flux density (right) for the full sample ($N=443$) and the sub-sample with robust PAH mass fractions ($N=216$) defined as those with {\qpah} error $<0.5\%$ or {\qpah} parameter constrained by $>3\sigma$ from the SED fitting (Section~\ref{sec:final-sample}). The sample is purely star-forming with AGN and quiescent galaxies removed (Section~\ref{sec:final-sample}).}
    \label{fig:sample}
\end{figure*}

\subsubsection{FRESCO data and redshifts}
The field of SMILES is fully covered by the FRESCO-GOODS-S F444W grism survey (PID 1895, \citealt{oesch23}), which results in the detection of Pa$\alpha$ at $z\sim 1.0-1.7$ and Pa$\beta$ at $z\sim 2.0-2.9$.

The details of the grism data reduction and extractions are in \cite{oesch23}. We extracted the grism spectra for 1,296 MIRI-selected galaxies ($3\sigma$ detected in F560W or F770W) at $z=0.5-3.5$ based on existing spectroscopic redshifts from the literature and photometric redshifts based on HST/ACS$+$JADES/NIRCam. The extractions follow the standard continuum$+$contamination removal procedure explained in \cite{oesch23} and \cite{kashino23}. The spectra were fit within $\Delta(z)= \pm 0.02 (1+z_{\rm spec})$ for the objects with existing spec-$z$; otherwise, within the lower and upper 68\% quantiles of their photo-$z$. The redshift probability distributions were then calculated mainly based on the detections of Pa$\alpha$ at $z=1.07-1.66$, Pa$\beta$ at $z=2.03-2.88$, and/or He{\sc i}$\lambda1083$, [S{\sc iii}], [S{\sc ii}], and higher-order Paschen lines. We visually inspected all ($N=$1296 objects) of the 1D extractions, the line fits, and accordingly evaluated the quality of the redshift estimations.
The final sample statistics are: 30\% ($N=386$) have robust fits and redshifts, out of which 56\% had prior spectroscopic redshifts; 50\% ($N=653$) of the fits are unreliable (e.g. no line, incomplete spectral coverage, contamination from other sources, etc.); and 20\% ($N=257$) are flagged as undetermined quality (noisy spectra, a weak line, or multiple lines that give different solutions).
In total, 57\% of galaxies in the MIRI-selected sample ($N=743$ galaxies) have either a FRESCO redshift or another spectroscopic redshift (3DHST and other literature).

\subsection{Final sample} \label{sec:final-sample}
We started from the MIRI-selected sample (3$\sigma$ detected in F560W or F770W) and used the following criteria to define our final sample: 
\begin{enumerate}
    \item $z=0.7-2.0$ to ensure adequate PAH coverage by MIRI data.
    \item $3\sigma$ detection in at least one MIRI band longwards of 15{\um}.
    \item The removal of quiescent galaxies based on their rest-frame $U-V$ and $V-J$ colours (following Equation 3 in \citealt{schreiber15}). For a cleaner selection, we also removed galaxies with specific SFR (sSFR, SFR/$M_*$ from SED modelling, see Section~\ref{sec:sed-fitting}) $<10^{-10}$\,yr$^{-1}$.
    \item The removal of objects with unusual or incorrect photometry from visual inspections of the SED fits.
\end{enumerate}

This selection results in 443 galaxies (`full' sample). The stellar mass and SFR distribution is shown in Figure~\ref{fig:MS-SEDs}. Specifically for the PAH analysis in this work, we further limited the sample to those with `robust' PAH mass fraction ({\qpah}, see Section~\ref{sec:sed-fitting-dust}). This category is defined as galaxies with either $>3\sigma$ {\qpah} values or {\qpah} error of less than 0.5\%. The second condition allows the inclusion of near-zero PAH fractions that are still relatively robust.
The rest of the sample is denoted as `low confidence' {\qpah} throughout the paper. The robust {\qpah} sample has 216 objects. Figure~\ref{fig:sample} shows the distributions of redshift, F444W (tracing emission from old stellar populations), and F1500W (tracing PAH and dust emission) of the robust and full samples. The robust {\qpah} sample is skewed towards brighter galaxies (and relatively more massive and dusty), by definition. Even though this sub-sample is biased, the separation is necessary to have a sample with reliable PAH fraction estimates. However, we show the results and trends for the low-confidence sample as well as the robust sample in the paper. The properties of the robust {\qpah} sample are provided in an online repository\footnote{Online repository for supplementary data: \href{https://zenodo.org/records/12671075}{https://zenodo.org/records/12671075}} and a sample table is shown in Table~\ref{table:qpah_data}.

\begin{table*}
\caption{Properties of the robust {\qpah} sample (Section~\ref{sec:final-sample}). Full table is published in an online repository: \href{https://zenodo.org/records/12671075}{https://zenodo.org/records/12671075}. Table contains: source ID from SMILES internal v0.3 catalog, redshift, PAH fraction and error, stellar mass and error in {\Msun}, metallicity (12+log(O/H)) and error calculated from the fundamental metallicity relation (FMR; as plotted in Figures~\ref{fig:qpah-metal} and \ref{fig:qpah_comparison}) and from the mass metallicity relation (MZR; as plotted in Figure~\ref{fig:pah-mass}).
}
\label{table:qpah_data} 
\centering
\begin{tabular}{cccccccccc} 
\hline\hline
ID& $z$& {\qpah}& {\qpah} error & $M_*/${\Msun} & $M_*/${\Msun} error & ${\rm [O/H]}_{\rm FMR}$& ${\rm [O/H]}_{\rm FMR}$ error & ${\rm [O/H]}_{\rm MZR}$& ${\rm [O/H}]_{\rm MZR}$ error\\
\hline 
28 & 1.5534 & 1.17 & 0.30 & 1.12e+10 & 7.89e+09 & 8.62 & 0.14 & 8.39 & 0.26 \\ 
30 & 1.5617 & 3.12 & 0.91 & 2.82e+09 & 8.11e+08 & 8.25 & 0.07 & 8.22 & 0.11 \\ 
{...} 
\end{tabular}
\end{table*}

\subsection{Spectral energy distribution fitting}\label{sec:sed-fitting}
We adopted the {\sc PROSPECTOR} SED fitting code \citep{johnson21} to fit the UV to mid-IR photometry. PROSPECTOR is a highly flexible SED fitting code that is based on Bayesian forward modelling and Monte Carlo sampling of the parameter space using FSPS stellar populations \citep{conroy13}. We have adopted the `current' or `surviving' stellar mass throughout this paper, which takes into account losses throughout the evolution of stars in the galaxy via AGB winds, supernovae, etc. This is different from the mass of stars that are formed in the lifetime of the galaxy. In our sample, the ratio of the surviving to the formed stellar mass is on average 0.69 with a scatter of 0.04 (for a Chabrier IMF). Below, we first discuss the assumptions on modelling the dust emission (Section~\ref{sec:sed-fitting-dust}), then discuss the validity of the derived PAH fractions (Section~\ref{sec:qpah-methodology}). Finally, in Section~\ref{sec:sed-fitting-other}, we discuss the remaining assumptions in the SED fitting.
We note that the main results of this paper (i.e. PAH and dust emission) are mainly constrained by the assumed dust models \citep{draineli07}, and therefore using a different SED fitting code is not expected to significantly change the results. 

\subsubsection{Dust emission assumptions} \label{sec:sed-fitting-dust}
The dust emission models in PROSPECTOR are based on the models of \cite{draineli07}. These models assume a mixture of amorphous silicate grains and carbonaceous grains with a distribution of grain sizes chosen to reproduce the wavelength dependence of the Milky Way (MW) extinction curve, while the silicate and carbonaceous content was constrained by the ISM gas phase depletion observations. The radiation field heating the dust is assumed to have a fixed shape (spectrum) of the local interstellar radiation field \citep{mathis83}, scaled by a dimensionless factor, $U$. Three parameters of these models are free in the PROSPECTOR framework: {\qpah}, the fraction of the grain mass contributed by PAHs containing fewer than $10^3$ carbon atoms, U$_{\mathrm{min}}$, the intensity of the diffuse ISM radiation field heating the dust, and $\gamma$, the fraction of dust mass exposed to the power-law distribution of starlight radiation intensities between U$_{\rm min}$ and U$_{\rm max}$ (($1-\gamma$) is the fraction of dust mass exposed to U$_{\rm min}$).  

In the absence of far-IR constraints, the IR parameters are constrained based on the energy balance assumption between the UV-near-IR dust attenuated luminosity and the re-emitted IR luminosity. \cite{leja17} experimented with the output with and without far-IR data on a large sample of local galaxies. They concluded that in the absence of far-IR constraints, it is critical to have tight priors on the shape of the far-IR emission to avoid unrealistically large uncertainties and potential systematic biases in SFR, $L_{\rm IR}$, and dust attenuation. Following their recommendation, we adopted a flat prior of $0.1<U_{\rm min}<15$ and $0.0<\gamma<0.15$. For the {\qpah} parameter we adopted a flat prior of $0.4<{\rm q_{PAH}}<4.6$, as this is the range of {\qpah} that the \cite{draineli07} models are specifically constrained for. We derived {\qpah} for a sub-sample of the galaxies with ALMA 1.1 and 3\,mm observations and find no systematic offset between the derived values including and excluding the ALMA data, as is explained in Section~\ref{sec:qpah-methodology}. 

\begin{figure*}[h!]
\centering    \includegraphics[width=.8\textwidth]{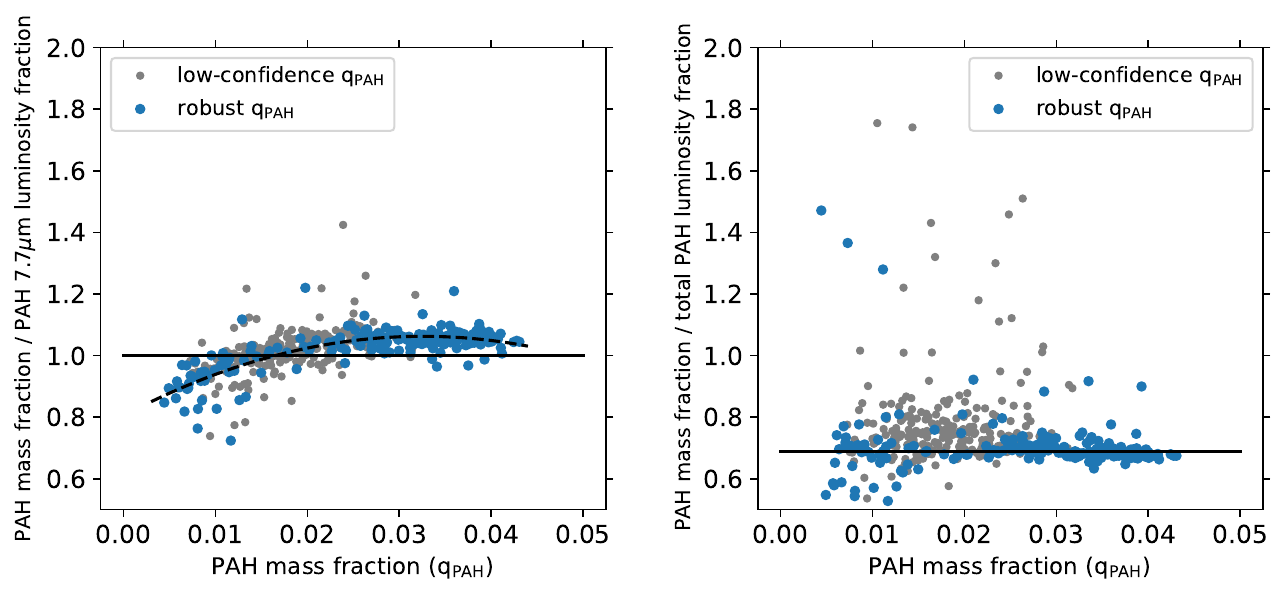}
    \caption{Ratio of the PAH-to-dust mass fraction to the PAH-to-total IR luminosity fraction versus the PAH mass fraction in our sample. Both quantities are derived from the \cite{draineli07} models (Section~\ref{sec:sed-fitting}). Robust {\qpah} sample shows galaxies with {\qpah} with $>3\sigma$ significance or less than 0.5\% error (Section~\ref{sec:final-sample}).
    Left: PAH luminosity fraction is defined as the ratio of the PAH 7.7\,{\um} feature luminosity to total IR luminosity. Horizontal line is the unity line. The robust sample is fit with a polynomial function shown with a dashed curve (Equation~\ref{eq:pah-mass-lum}). Right: PAH luminosity is the sum of the 3.3, 6.2, 7.7, 11.2, and 17\,{\um} PAH features luminosities (Section~\ref{sec:sed-fitting-dust}). Horizontal line shows the median of the robust sample (0.68).
    We note that {\qpah} is not in percentage units, unlike the other figures in this paper.}
    \label{fig:pah-mass-lum}
\end{figure*}
We note that the fraction of dust mass in PAHs used in this work, denoted as {\qpah}, should be distinguished from the fraction of total IR luminosity in PAH bands. The two quantities are related to each other but not with a constant value, as the light-to-mass ratio of dust is not a fixed value in the \cite{draineli07} models (it is a function of radiation field intensity for the dust that produces the IR continuum and for PAHs, it depends on the stochastic heating).
In Figure~\ref{fig:pah-mass-lum}, we compare the two quantities for our sample. The PAH mass fraction is compared to the PAH luminosity fraction defined as either a ratio of PAH 7.7\,{\um} feature ($F_{7.7}$, left panel) or a total PAH luminosity (right panel) to total IR luminosity (integrated luminosity from 3 to 1100\,{\um}). $F_{7.7}$ is calculated as the integrated PAH flux ($\int (F_{\lambda}-F^{\rm{continuum}}_{\lambda})~{\rm d}\lambda$) between 6.9 and 9.7{\um} and is corrected for the continuum emission by assuming zero feature strength on either side of the feature at 6.9 and 9.7{\um}, following the recipe of \cite{draine21}. The total PAH luminosity is the sum of the luminosity of the 3.3, 6.2, 7.7, 11.2, and 17\,{\um} features, measured in the same way as the 7.7\,{\um} luminosity within the windows defined by \cite{draine21}. As is shown in the figure, the PAH mass fraction can be converted to the total PAH luminosity fraction by a factor of 0.68 (the median value of {\qpah} to total PAH luminosity fraction); however, there is a large scatter around this median value. The {\qpah} to PAH 7.7\,{\um} luminosity fraction is tighter, but the relation with {\qpah} is not linear. We fitted the relationship with a polynomial function:
\begin{equation}\label{eq:pah-mass-lum}
    \frac{{\rm q_{PAH}}}{F_{7.7}/F_{\rm TIR}} = -0.024\times{\rm q_{PAH}}^2 + 0.158\times{\rm q_{PAH}} + 0.805,
\end{equation}
where $F_{7.7}/F_{\rm TIR}$ is the PAH 7.7\,{\um} to total IR luminosity fraction, and {\qpah} is the fraction of PAH to dust mass. The {\qpah} values presented in this work can be converted to the PAH 7.7\,{\um} luminosity fraction using this equation, if needed.

\begin{figure*}[h!]
\centering    \includegraphics[width=.8\textwidth]{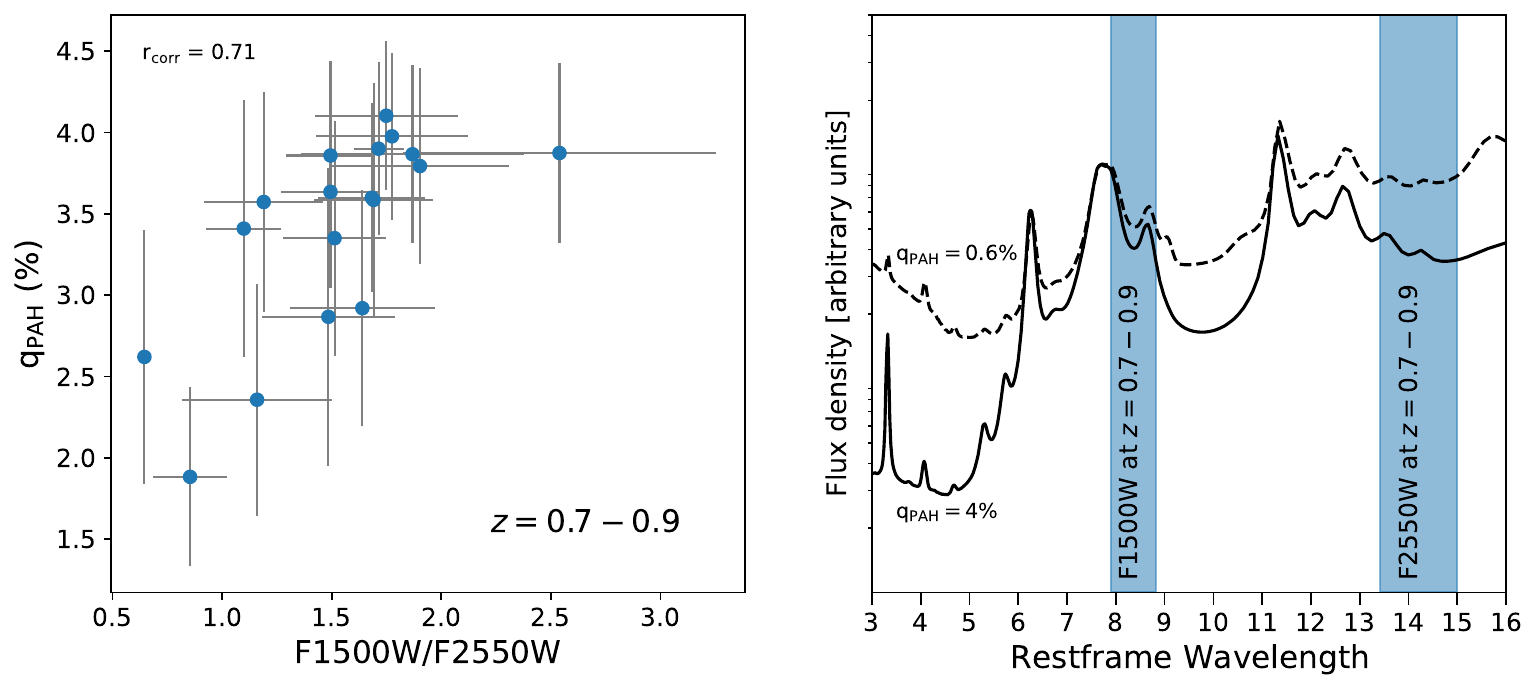}
    \caption{Left panel: PAH-to-dust mass fraction ({\qpah}) versus 15{\um} (F1500W) to 25.5{\um} (F2550W) flux ratios for a sample of $z=0.7-0.9$ galaxies with significant ($>3\sigma$) detections in F1500W and F2550W. {\qpah} is derived from PROSPECTOR (the \citealt{draineli07} models), while the flux ratio is an empirical tracer of PAH to warm dust emission, as is shown in the right panel. The significant linear correlation of the two quantities (Pearson correlation coefficient of 0.71 with p-value = 0.001) indicates the validity of the estimated {\qpah} values.
    Right panel: Two mid-IR spectra with a high PAH fraction (solid) and low PAH fraction (dashed), overlaid with the wavelength coverage of the F1500W and F2550W filters at $z=0.7-0.9$. In this redshift range, F1500W and F2550W filters trace the PAH dominated and non-PAH-dominated dust emission in star-forming galaxies, respectively. Therefore, their ratio is expected to correlate with PAH fraction. }
    \label{fig:qpah-test}
\end{figure*}

\subsubsection{Robustness of {\qpah} inferred from photometry} \label{sec:qpah-methodology}

The majority of our sample does not have IR detections beyond MIRI. Therefore, the PAH mass fractions were inferred solely based on UV to mid-IR (MIRI) data. As a validity check for the derived {\qpah} parameters, we performed two tests, as follows.

We compare the {\qpah} values for a sample of 18 galaxies at $z=0.7-0.9$ that have significant ($>3\sigma$) detections at 15 and 25.5{\um}. As is shown in the right panel of Figure~\ref{fig:qpah-test}, $z=0.9$ is the highest redshift that our longest-wavelength band (F2550W) traces the part of the SED that is dominated by the continuum dust emission, and not that of PAHs. At $z=0.7-0.9$, the F1500W filter traces the $7-8${\um} PAH complex. Therefore, the ratio of F1500W to F2550W fluxes traces the PAH to warm dust emission, which, in theory, is expected to scale with the PAH-to-dust mass fraction (i.e. {\qpah}). As is shown in the left panel of Figure~\ref{fig:qpah-test}, the derived {\qpah} values from the SED fits are highly correlated with the F1500W to F2550W flux ratio, with a Pearson correlation coefficient of 0.71 (p-value $\ll1$).

\begin{figure}[h!]
\centering    \includegraphics[width=.45\textwidth]{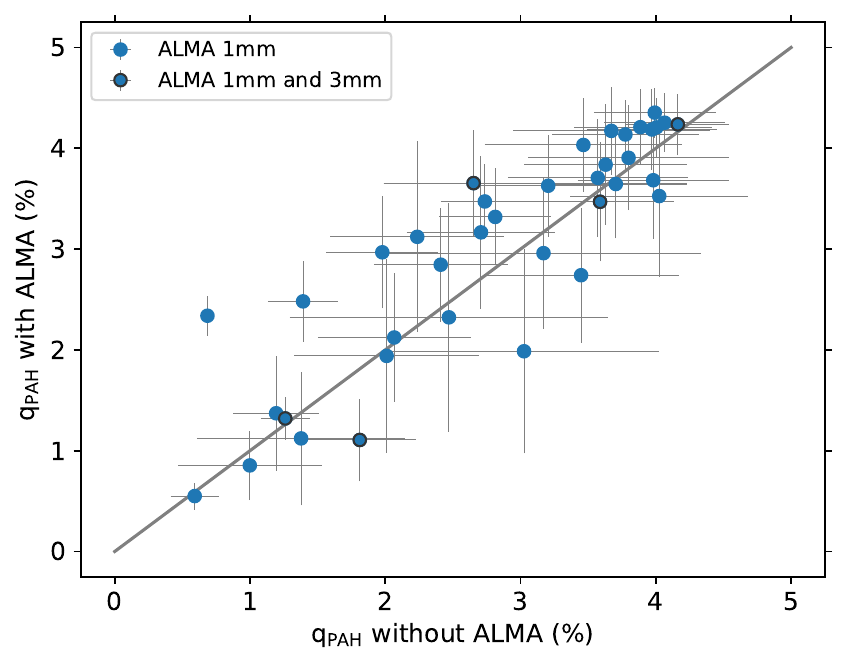}
    \caption{PAH to total dust mass fraction for the 38 galaxies with ALMA 1.1\,mm detection \citep[from the ASPECS survey;][]{walter16}. Five of them also have 3\,mm detections. The axes show the derived {\qpah} including and excluding the ALMA data points.
    Median of the {\qpah} values without ALMA to those with ALMA data is 0.95, completely within the {\qpah} uncertainty, indicating the {\qpah} values inferred based on UV-to-MIRI data alone are robust. 
    }
    \label{fig:qpah-alma}
\end{figure}
As {\qpah} depends on both the PAH mass and the mass of the other dust grains, the parameter may be uncertain in the absence of far-IR data.
There are 38 SMILES galaxies with ALMA 1.1\,mm (band 6) detection from one of the deepest ALMA extragalactic surveys, ASPECS \citep{walter16}. Five of these also have 3\,mm detections (band 3). We compare the derived {\qpah} of this sample including and excluding ALMA data in Figure~\ref{fig:qpah-alma}. There is no systematic difference in {\qpah} within the uncertainties. Given the luminosity range of these galaxies, this is in agreement with previous studies that showed PAHs as good tracers of IR luminosity in relatively bright (and non-ULIRG) galaxies \citep{reddy12a,shipley16}. Therefore, we conclude that our derived {\qpah} values are robust even without far-IR data under our fitting assumptions. 

\subsubsection{Other spectral energy distribution assumptions} \label{sec:sed-fitting-other}
We assumed a delayed-$\tau$ star formation history with a flat age prior between 1\,Myr to the age of the Universe at the redshift of the galaxy. Stellar metallicity is free with a flat prior between $\log(Z/Z_{\odot}) = -2.5-0.19$. 
For dust attenuation, we adopted a flexible attenuation curve that has its UV slope as a free parameter as parameterised in \cite{kriek13}, with a flat prior between $-0.6$ to 0.3 for the multiplicative coefficient of the slope of the \cite{calzetti00} curve. We included AGN emission, even though the AGN fraction of our sample is very low, by the definition of the sample (median of 0.05\% AGN-to-bolometric stellar luminosity fraction). We assumed a nebular emission model with flat priors in logarithmic space of gas-phase metallicity at $-2.0$ to 0.5 (relative to solar metallicity) and ionisation parameter at $-4$ to $-1$. We included IGM absorption \citep{madau95} with the optical depth scaling factor as a free parameter with a Gaussian prior centred at 1.0 and $\sigma=0.3$.

Redshift was fixed if the spectroscopic redshift was available from various non-JWST spectroscopic campaigns (compilations of \citealt{kodra23}, \citealt{bacon23}) or JWST JADES NIRSpec \citep{bunker23} and FRESCO NIRCam grism \citep{oesch23} surveys. In the absence of spec-$z$, the redshift parameter was left free with a a clipped-normal prior distribution centred at the photometric redshift and limits of the photo-$z$ uncertainty. The photo-$z$ was taken from the JADES photometric redshift catalogue \citep{hainline23}, with typical uncertainties of $\Delta(z)/z = 0.04-0.1$ (median of 0.06).

\subsection{MIRI sizes} \label{sec:method-size}
Dust sizes and surface areas were estimated from MIRI data as a proxy for where the bulk of stars are forming in galaxies. These measurements were later used to calculate SFR surface densities. 
We calculated dust sizes from MIRI F1500W images with 0.5'' resolution, tracing rest-frame $5-9$\,{\um}. The choice of the F1500W filter to measure dust sizes was made as this filter provides a good compromise between wavelength and spatial resolution.

We used WebbPSF \citep{perrin14} to model the F1500W PSF. In order to estimate sizes, we used GALFIT \citep{peng02,peng10} to fit the spatial profile of each source.  Each one was modelled twice: once with a point-source model and once with a single-Sersic model.
For a complete morphological analysis, more complicated models are required but for our current purposes these simple models provide reasonable sizes for the majority of the sample.
Galaxies that fit better (smaller reduced $\chi^2$) with a S\'{e}rsic model than the point-source model are considered spatially resolved. Sources that are best fit with the point source model or with a S\'{e}rsic profile but whose half-light radius is smaller than the 50\% encircled energy radius for the appropriate filter are considered unresolved. For the unresolved sources, we used the 50\% encircled energy size of the PSF as an upper limit on their effective radius. We visually inspected all fits and removed the bad fits (e.g. if the source had signs of merger or nearby bright companions). This results in 105 F1500W good fits for extended objects in our sample.

\begin{figure*}
        \centering
        \includegraphics[width=.48\textwidth]{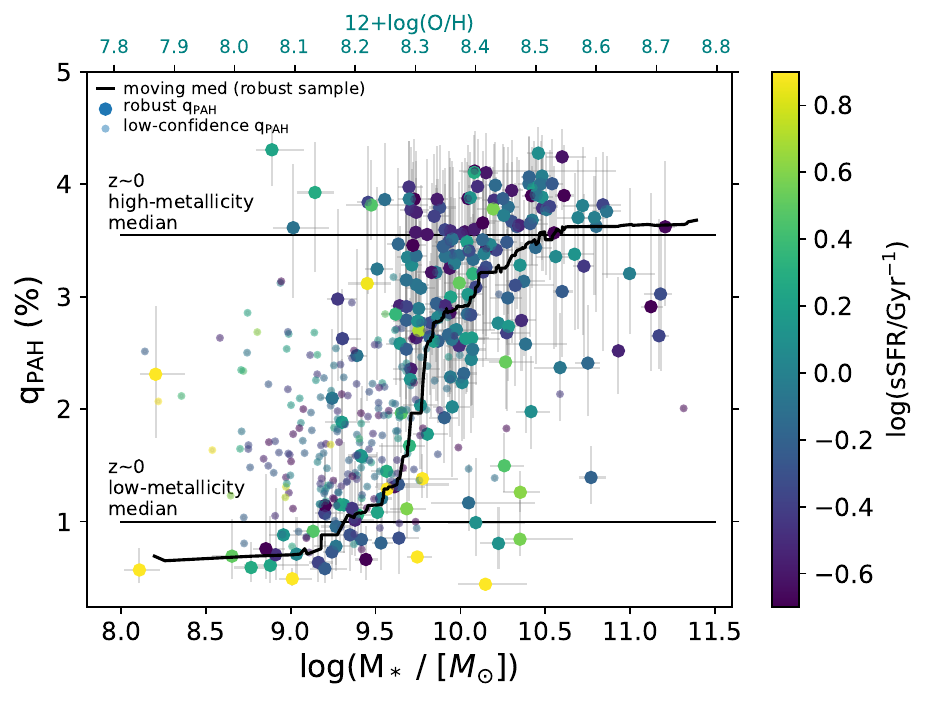} \quad
        \includegraphics[width=.49\textwidth]{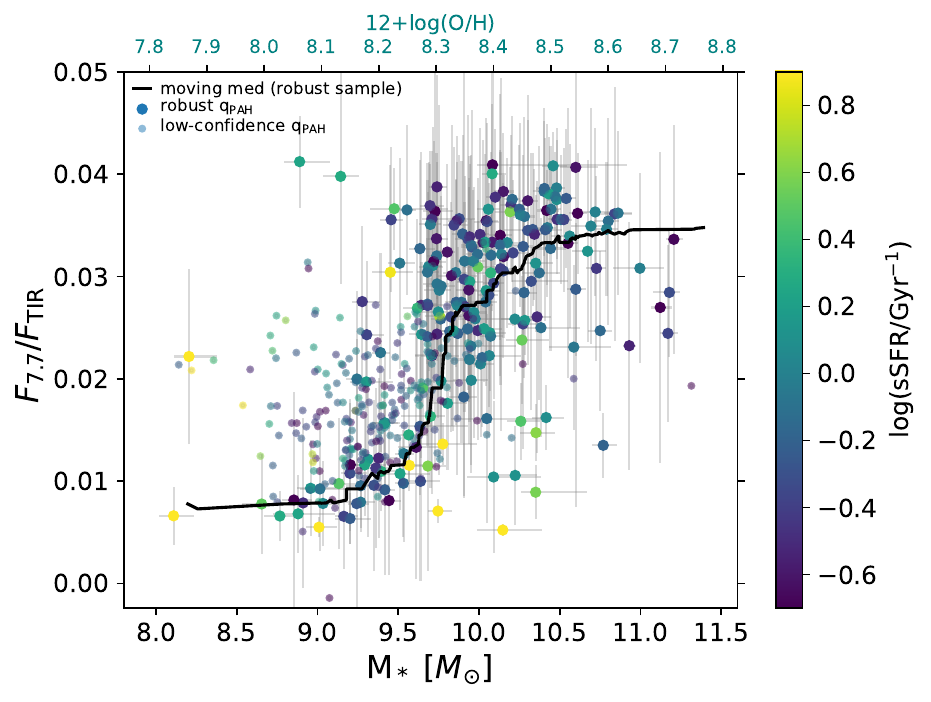}
    \caption{PAH mass fraction (left) and luminosity fraction (right) as a function of stellar mass. Moving median (black curve) is shown for the robust {\qpah} sample (see Section~\ref{sec:final-sample} for the definition of samples), and all galaxies are colour-coded by sSFR. In the left panel, horizontal lines show the median PAH mass fractions of $z\sim 0$ galaxies with {\logoh}$<8.1$ and $>8.1$ from \citet{draine07b}. On the top x-axes, we also show the estimated metallicities from the O3N2-based mass-metallicity relation (MZR) of \cite{topping21} at $z\sim 1.5$, and assuming the \cite{pp04} calibration. The $z\sim 0.7-2$ sample has a very similar {\qpah}-metallicity behaviour to that inferred at $z\sim 0$. 
    Properties of the robust sample are in Table~\ref{table:qpah_data}.
    }
    \label{fig:pah-mass}
\end{figure*}

\begin{figure*}
        \centering
        \includegraphics[width=\columnwidth]{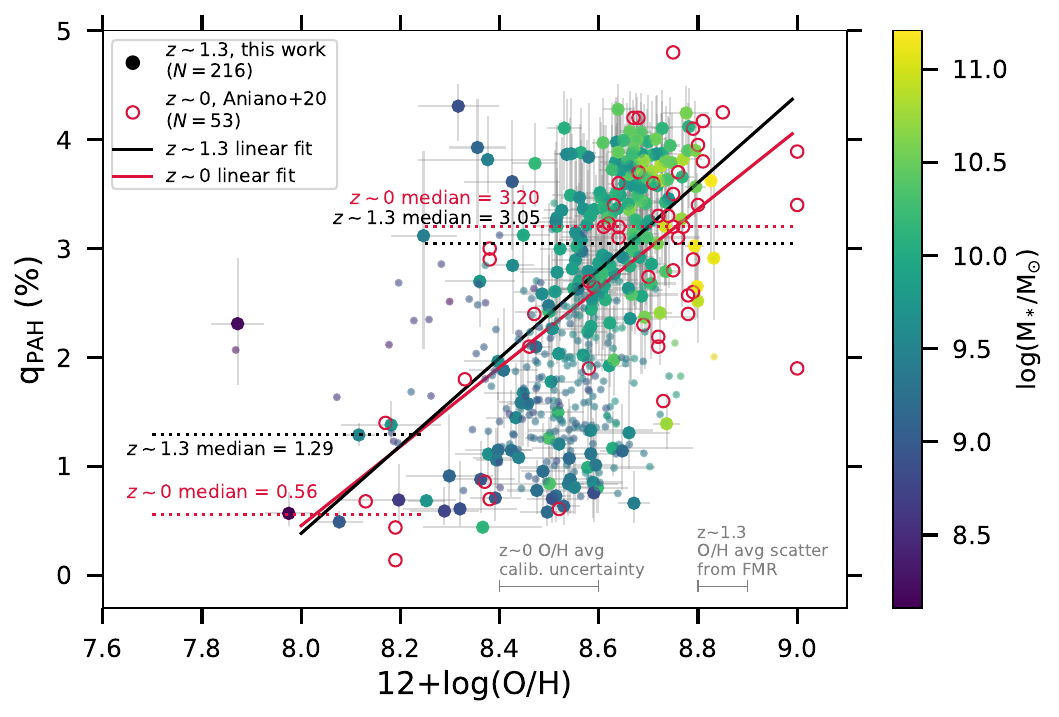}
    \caption{PAH mass fraction versus metallicity. Metallicity was derived from the FMR of \cite{sanders21}. The robust and low-confidence {\qpah} samples (Section~\ref{sec:final-sample}) are shown with larger and smaller circles, respectively. We provide the values of the robust sample in Table~\ref{table:qpah_data}.
    The $z\sim 0$ data from \cite{aniano20} are shown in red. Following the analysis of \cite{aniano20}, we fit both samples with a linear function, shown with solid lines (slopes of $3.63\pm0.71$ and $4.04\pm0.51$ and intercepts of $-28.60\pm6.14$ and $-31.94\pm4.38$ for the $z\sim 0$ and 1.3 sample, respectively). 
    The medians of the two samples below and above {\logoh}$=8.25$ (dotted lines) are also shown. The two samples follow very similar trends.
    The average scatter in {\logoh} from FMR (0.05\,dex, see text) and the strong-line metallicity calibration uncertainty for the $z\sim 0$ sample (0.2\,dex; \citealt{aniano20}) are shown in the bottom right. The error bars on the $z\sim 1.3$ data are the metallicity measurement uncertainties from the SFR and mass uncertainties.
    The linear fits (solid lines) have slopes of 3.6 and 3.4 for the $z\sim 0$ and 1.3 samples, respectively. 
    }
    \label{fig:qpah-metal}
\end{figure*}

\begin{figure*}
        \centering
        \includegraphics[width=.8\textwidth]{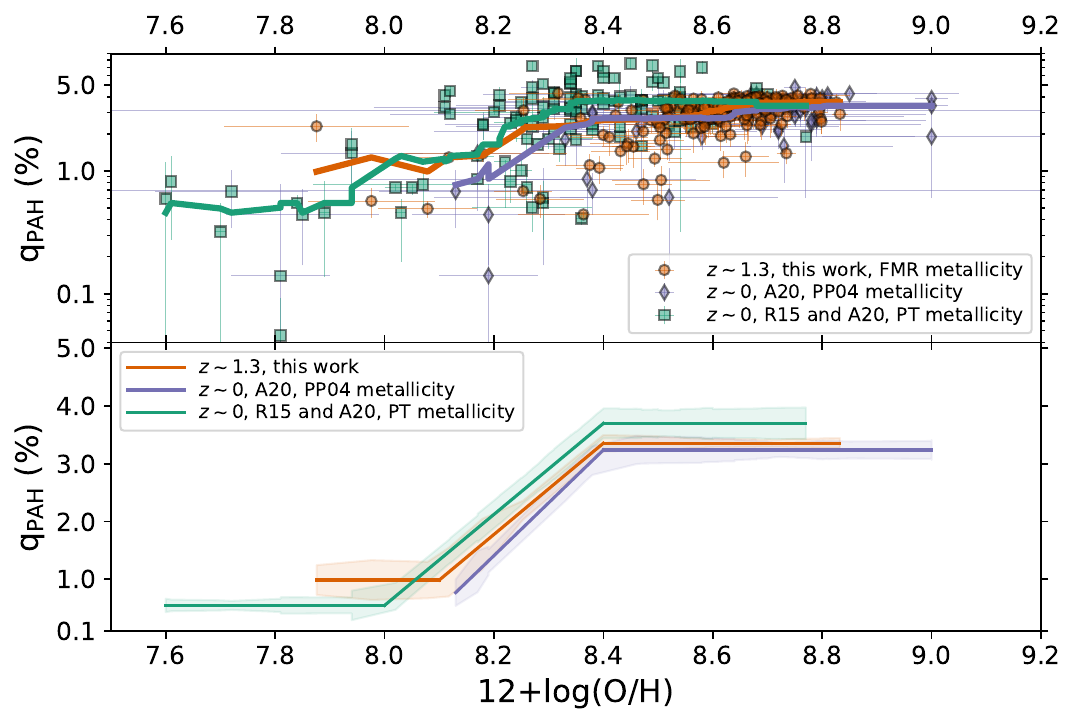} 
    \caption{PAH mass fraction as a function of metallicity at $z\sim 1.3$ (this work) and $z\sim 0$ \citep{aniano20,remyruyer15}. Here, we only include the robust {\qpah} sample at $z\sim 0.7-2.0$. 
    Top: The $z\sim 1.3$ data presented in this work is shown with orange circles. The orange line shows the moving median. Metallicity of the $z\sim 1.3$ sample is estimated using the FMR relation of \cite{sanders21}. 
    The $z\sim 0$ data is shown for the PP04 metallicity calibrations of \cite{aniano20} (purple diamonds, and purple line for the moving median) and the PT metallicity calibrations of \cite{aniano20} and \cite{remyruyer15} (green squares, and the green line for the moving median). The \cite{remyruyer15} data extend the $z\sim 0$ sample to much lower metallicities. The two metallicity calibrations are systematically different (on average $\sim 0.2$\,dex but can be up to 0.4\,dex); therefore, we show both for completeness. 
    Bottom: Three-step fits (constant$+$linear$+$constant) to the moving median trends in the top panel. The shaded regions show the error in the mean in each consecutive bin of the moving median. Despite the uncertainties in metallicities, the overall {\qpah}-metallicity behaviour of the $z\sim 0$ and $z\sim 1.3$ samples are similar within the scatter.
    }
    \label{fig:qpah_comparison}
\end{figure*}

\begin{figure*}
        \centering
        \includegraphics[width=.8\textwidth]{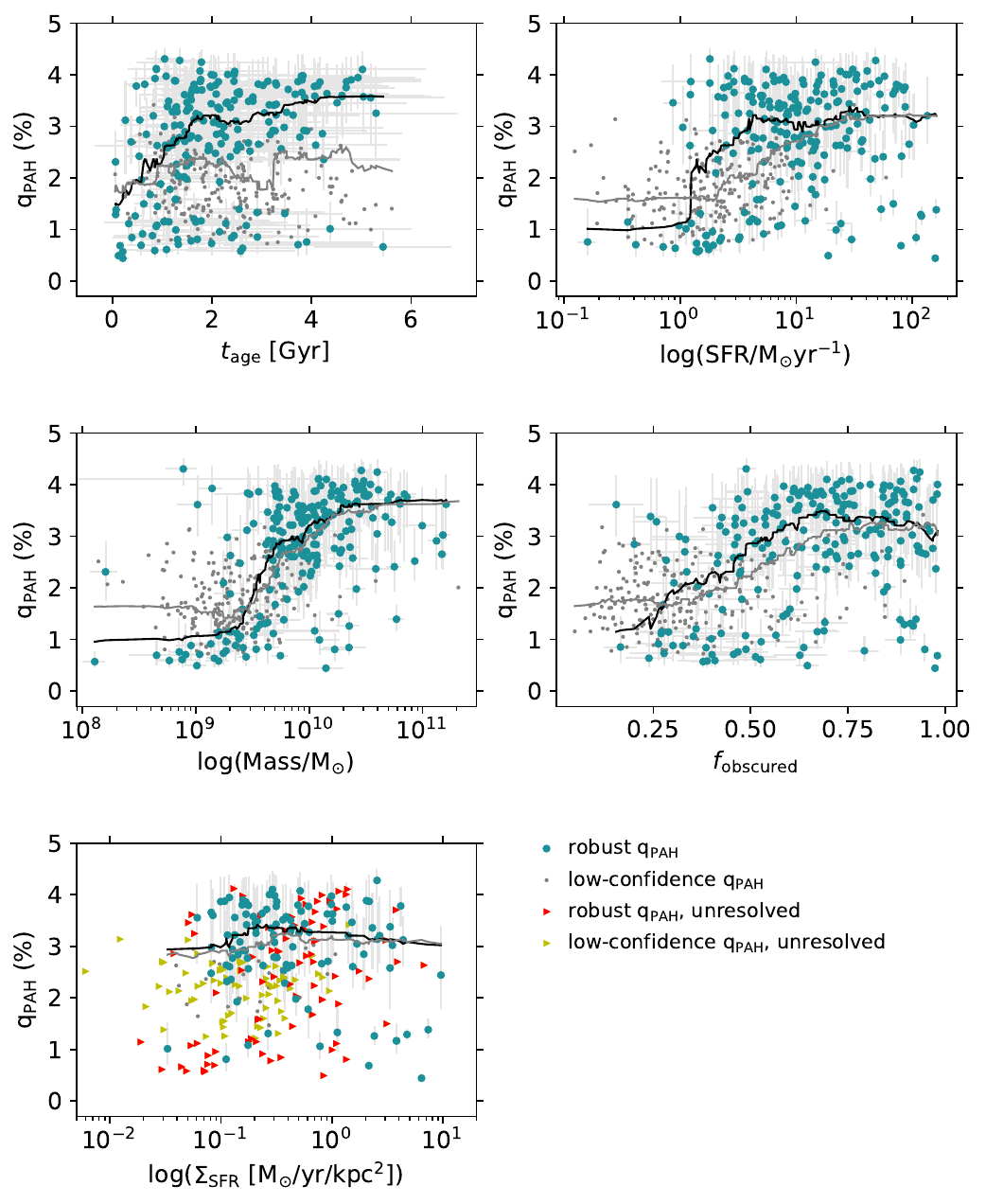} 
    \caption{PAH mass fraction as a function of galaxy parameters. From top-left to bottom-right: age of the galaxy, SFR, stellar mass, obscured luminosity fraction, SFR surface density (using F1500W sizes).
    The robust sample is shown with a teal colour and the low-confidence sample with smaller grey symbols (see Section~\ref{sec:final-sample} for sample definitions). The moving medians of the full (robust$+$low-confidence) and robust samples are shown as grey and black curves, respectively.
    In the bottom row, the unresolved sources in F1500W images are shown with triangles as lower limits on SFR surface density.
    Among the parameters show, {\qpah} has the strongest correlations with mass, age, and obscured luminosity fraction. 
    }
    \label{fig:qpah_galpar}
\end{figure*}

\section{Evolution of the polycyclic aromatic hydrocarbon fraction}
\label{sec:pahs}
Figure~\ref{fig:MS-SEDs} shows the full and the {\em robust} ({\qpah} determined at $>3\sigma$) samples in the SFR-$M_*$ (star-forming main-sequence) diagram. The colours indicate the inferred {\qpah} for the robust sample, which ranges from $\sim 0.5-4.5$\%. The galaxies are representative of `main sequence' galaxies at these redshifts \citep{leja22} and span a wide range in sSFR. Examples of the SED fits for three objects that are above, on, and below the main-sequence are shown in the right panels. In the following sections, we investigate the relationship between PAH fractions and galaxy parameters, and its evolution from $z\sim 2$ to $z\sim 0$.

\subsection{Polycyclic aromatic hydrocarbon fraction versus stellar mass}
In Figure~\ref{fig:pah-mass}, we show the relationship between stellar mass and {\qpah} (left panel) and the 7.7\,{\um} PAH luminosity fraction relative to total IR luminosity emission ($F_{7.7}/F_{\rm{TIR}}$; right panel). $F_{7.7}$ is the integrated PAH flux ($\int (F_{\lambda}-F^{\rm{continuum}}_{\lambda})~{\rm d}\lambda$) between 6.9 and 9.7{\um} and is corrected for the continuum emission, following the recipe of \cite{draine21}, and the total IR (TIR) power is the integrated emission ($\int F_{\lambda}~d\lambda$) from 3 to 1100{\um} of the best-fit SED model.

The PAH `strength', defined as either the PAH mass or emission fraction, is positively correlated with stellar mass with a correlation coefficient of $\rho_{\rm Pearson} = 0.5$ for the robust {\qpah} sample and 0.6 for the full sample (with p-values of $\ll 1$). We also estimate the gas-phase metallicity ({\logoh}) based on stellar masses and the $z\sim 1.5$ O3N2-based mass-metallicity relation (MZR) of \cite{topping21} and the \cite{pp04} calibration, shown on the top axis of Figure~\ref{fig:pah-mass}.
The median trend of {\qpah} with the MZR-estimated metallicities closely resembles the local galaxies trend from \cite{draine07b} (the horizontal lines in Figure~\ref{fig:pah-mass}-left). The two {\qpah} median values of the local galaxies were reported at metallicities below and above {\logoh}$=8.1$. The rapid change in {\qpah} happens at $\sim 0.1-0.2$\,dex higher oxygen abundances in our $z\sim 1.3$ sample compared to that at $z\sim 0$. However, given the uncertainties in metallicity calibrations (see also the next section), we conclude that the behaviour of the $z\sim 1.3$ sample is consistent with that at $z\sim 0$.

There is a large scatter in {\qpah} at intermediate masses, such that in the mass range of $\log(M_*/M_{\odot})\sim 9.5-10.5$, there are galaxies with {\qpah} of $<0.5\%$ to $>4\%$. The majority of these galaxies seem to have young ages and high specific SFRs (sSFR).
To better understand the physical process driving the scatter, we also calculate partial correlation coefficients between {\qpah}, $M_*$, and a third galaxy parameter, including sSFR, SFR, age, optical depth, obscured SFR fraction, and SFR surface density. However, we do not find a significant secondary dependence on any of these parameters in the {\qpah}-$M_*$ correlation of the robust sample. If we limit the sample to only {\qpah} values with $>3\sigma$ significance, the specific SFR (sSFR) and age reduce the {\qpah}-$M_*$ correlation coefficient of 0.4 to a partial coefficient of 0.25, indicating a potential secondary influence on the {\qpah}-$M_*$ correlation, such that at a given stellar mass, galaxies with higher sSFRs and younger ages have lower PAH strengths. 

\subsection{Polycyclic aromatic hydrocarbon fraction versus metallicity from $z\sim 0$ to 2}

The relation between the PAH fraction and metallicity has been extensively studied in the local Universe \citep{marble10,engelbracht05,madden06,smith07,galliano08,draine07b,remyruyer15,aniano20,li20,chastenet23} and also at cosmic noon \citep[][using \textit{Spitzer} data]{shivaei17}. To investigate this relation with the deeper MIRI data at $z>0$, we estimate metallicities from the fundamental metallicity relation (FMR) between stellar mass, SFR, and metallicity. The FMR is shown to be established and invariant up to $z\sim 3$ \citep{cresci19, sanders21}. \cite{sanders21} estimated the scatter in {\logoh} from FMR to be 0.06\,dex at $z\sim 2.3$, and 0.04\,dex at $z\sim 0$. This is typically smaller (or at most comparable) to the systematic uncertainty in strong-line metallicity calibrations (e.g. $\sim 0.2$\,dex average systematic uncertainty between the \citet{pp04} and the \citet{pt05} calibrations in \cite{aniano20}, see their Figure 1). We adopt the FMR presented in \cite{sanders21} to estimate metallicities from our mass and SFR values. The adopted FMR is in agreement with those derived from direct-method metallicities \citep{andrews13} and also with the best-fit for local galaxies in \cite{curti20a}. 
We adopt a Monte Carlo simulation by randomly perturbing stellar mass and SFR within their uncertainties for 500 times, and derive the metallicities according to the FMR relation. The final metallicity and uncertainty of each object is the median and standard deviation of the 500 realisations. 
The average measurement uncertainty of the inferred {\logoh} is 0.04\,dex, comparable to the FMR scatter in {\logoh}.

We limit the $z\sim0$ comparison samples to two studies that adopted, similar to this study, the \cite{draineli07} models to derive {\qpah} and have homogeneous metallicity calculations: \cite{aniano20} (hereafter, A20) and \cite{remyruyer15} (hereafter, R15). A20 used two strong-line metallicity calibrations for the KINGFISH sample: 1) the `PT' or \cite{pt05} method, adopted from \cite{moustakas10}, and 2) the `PP04' or \cite{pp04} based on [N{\sc ii}]/H$\alpha$. They showed that the two metallicities can differ by as much as 0.4\,dex (on average $\sim 0.2$\,dex), reflecting the large uncertainty in the calibrations. 
R15 incorporated the Dwarf Galaxy Survey (DGS; \citealt{madden13}) into the KINGFISH which extended the sample to lower metallicities. They adopted the PT metallicity calibration.
 
Figure~\ref{fig:qpah-metal} shows the {\qpah} trend with metallicity. There is a positive correlation (Pearson $\rho = 0.55$ with p-value $\ll1$) with a large scatter at high metallicities. The trend is very similar to the $z\sim 0$ trend from A20 (assuming their preferred PP04 metallicity\footnote{A20 and \cite{hunt16}, who used the same metallicities in their analysis, preferred the PP04 calibration as it shows tighter scaling relations with other calibrations and with {\qpah}.}). A20 approximated the {\qpah}-metallicity behaviour by both a linear function and a step function. The median values below and above their threshold metallicity, {\logoh}$\sim 8.25$, and the slopes of the linear fits ($3.6\pm0.7$ and $4.0\pm0.5$ at $z\sim 0$ and 1.3, respectively) for the two samples suggest a non-evolving PAH fraction-metallicity relation from $z\sim 0-2$.

In Figure~\ref{fig:qpah_comparison} (top panel), we compare our results with the $z\sim 0$ data over a wider range of metallicities by combining both the A20 and R15 data (using the PT metallicity). The {\qpah} values of R15 extend to larger values compared to A20 and ours, because they set a higher upper limit for the {\qpah} prior in their fittings. We also include A20 data for the PP04 metallicity, for completeness.
To interpret and compare the datasets, we adopt a moving (running) median to smooth the data without assuming a parametric functional form a priori. All three trends show a constant low ($\lesssim 1\%$) {\qpah} at low metallicities ({\logoh}$\lesssim 8.1-8.2$ or $\sim 0.3\,Z_{\odot}$ assuming {\logoh}$=8.69$ for Solar metallicity; \citealt{asplund09}), then a linear increase in {\qpah} with metallicity that ends with a plateau ({\qpah} $\sim 3.4\%$) at {\logoh}$\gtrsim 8.4$ ($\gtrsim 0.5\,Z_{\odot}$). We show the fits to the moving medians in the lower panel of Figure~\ref{fig:qpah_comparison}. The constant {\qpah} at high metallicities confirms the successful usage of PAH strengths as IR luminosity and obscured SFR indicators for moderately massive and metal-rich galaxies based on local calibrations \citep[e.g.][]{calzetti07,shipley16}. Our findings are also in general agreement with \cite{marble10} based on the $z\sim 0$ SINGS data, who found a linear relation between PAH luminosity fraction and metallicity. 
These observations show that, at least up to $z\sim 2$, metallicity is a good indicator of the ISM properties that affect the balance between the formation and destruction of PAHs in star-forming galaxies, and the responsible processes maintain a constant PAH abundances of $\sim 3.4$\% above a metallicity of $\sim 0.5\,Z_{\odot}$. 

Below $\sim 0.5\,Z_{\odot}$, we see that there is a linear decrease in the PAH fraction, which has a similar slope between the $z\sim 0$ and 1.3 samples. 
This universal PAH fraction-metallicity relation can potentially be calibrated to estimate the metallicities of metal-poor galaxies at high redshifts using IR imaging surveys alone.
The source of the decreasing PAH fraction with decreasing metallicity can be explained by the destruction of PAHs at low metallicities, either through radiative destruction in less shielded low-metallicity environments \citep{madden06,hunt10,narayanan23}, or supernova shocks \citep{seok14}. Some studies also explain it by insufficient production of PAHs due to a delayed injection into the ISM by AGB stars \citep{galliano08} or more efficient production at high metallicities through shattering of large grains \citep{seok14}. 
Future cosmological simulations with on-the-fly dust evolution simulations can shed light on the complex interplay of the formation and destruction processes of these grains in the ISM. 

\subsection{Polycyclic aromatic hydrocarbon fraction and other galaxy parameters}
In Figure~\ref{fig:qpah_galpar} we show {\qpah} as a function of other galaxy parameters: age, SFR, stellar mass, obscured luminosity fraction ($f_{\rm obscured}$), and SFR surface density ({\Ssfr}). Obscured luminosity fraction is calculated as the ratio of the dust-obscured UV (1550\AA) luminosity to the intrinsic UV luminosity (see Section~\ref{sec:obsc}). Sizes are calculated in the F1500W filter, tracing rest-frame 5 to 9\,{\um} dust emission, for galaxies with spatially resolved profiles (see Section~\ref{sec:method-size}). In the literature, {\Ssfr} is often calculated based on the area derived from a shorter-wavelength size measurement that is not a good proxy for where the bulk of the stars are forming. Here, surface area is based on dust sizes (from F1500W) that closely reflects the star-forming region, and hence, the resulting SFR surface densities are more reliable.

We show the moving medians for the robust (black curve) and the full sample (grey curve), which includes the robust sample, in Figure~\ref{fig:qpah_galpar}. In the robust sample, {\qpah} correlates with age, stellar mass, and obscured fraction. The {\qpah}-mass correlation is the strongest among these, and is likely a byproduct of the mass-metallicity relation (as was discussed in previous sections). The correlation with age can be expected as a stellar evolutionary effect if the AGB stars are the primary production channel of PAHs \citep{galliano08,li20} and, in fact, it is also seen at $z\sim 2$ from MIPS data \citep{shivaei17}. However, this correlation disappears when the full sample is considered. The correlation with obscured fraction is interesting as it may suggest PAH destruction due to reduced shielding by dust grains in systems with low obscured fractions.
We also note the lack of a significant correlation with SFR surface densities for the spatially resolved galaxies, which may suggest that the fraction of PAH-to-total dust mass is insensitive to the star formation intensity. Even including lower limits on SFR surface density for the unresolved sources (triangles in Figure~\ref{fig:qpah_galpar}) does not suggest a correlation with SFR surface density. Further analysis of this trend using size measurements from other filters that have higher resolution but still trace the dust emission will be done in a future work.

\begin{figure*}[h]
        \centering
        \includegraphics[width=.45\textwidth]{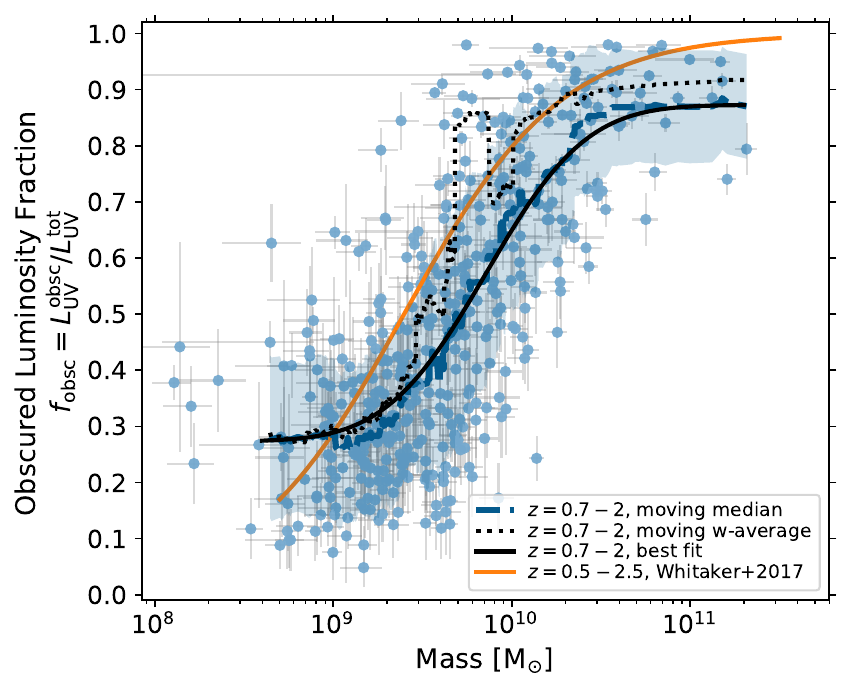} \quad
        \includegraphics[width=.45\textwidth]{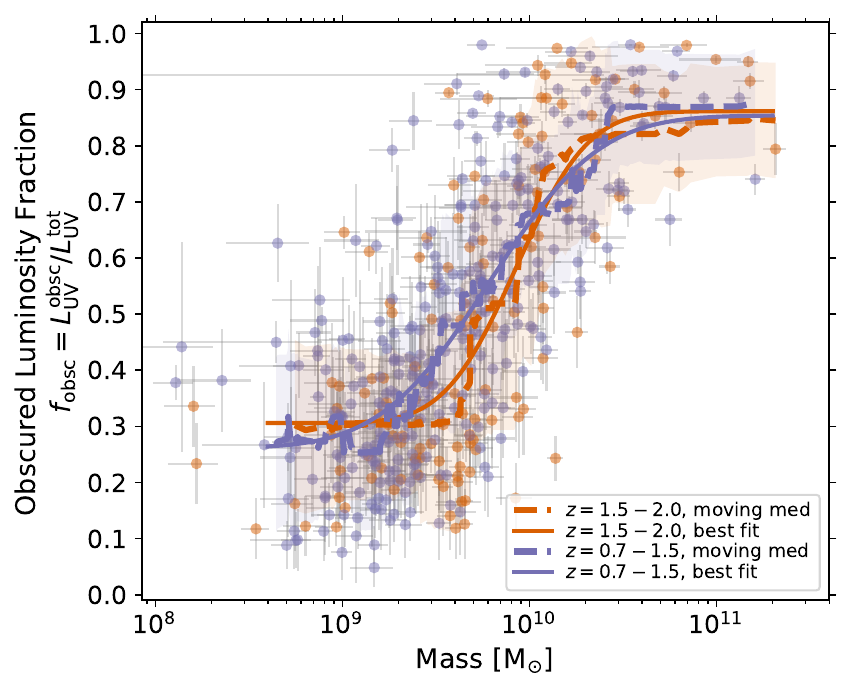}
    \caption{Obscured luminosity fraction as a function of stellar mass for the full sample (left) and separated by redshift (right). Obscured fraction is defined as the obscured UV to total (intrinsic) UV luminosity at 1550\,{\AA}, inferred from the best-fit UV-to-IR SEDs. 
    We show the moving median (dashed thick lines), moving weighted-average (dotted line) and best-fit curve (solid lines) to the samples at $M_*>4\times10^8${\Msun}, where the data is mass-complete.
    In the left panel, we also show the \citet{whitaker17} fit to $z= 0.5-2.5$ with an orange curve. The \citet{whitaker17} study was based on \textit{Spitzer}/MIPS data with a wider area but biased to more obscured systems. 
    Our weighted-average trend is also dominated by the heavily obscured systems, and therefore similar to the \citep{whitaker17} curve. However, the median trend that is more representative of the full population captured by the deeper MIRI data (relative to the MIPS data) is systematically lower. 
    Furthermore, as is shown in the right panel, there is no redshift evolution in this relation in our sample.
    }
    \label{fig:obsc_frac_w17}
\end{figure*}

\section{Dust-obscured luminosity fraction} \label{sec:obsc}

The largest uncertainty in measuring SFR is the fraction of dust-obscured star formation. The most reliable and direct method of accounting for the dust-obscured fraction is observing the dust thermal emission in IR. With \textit{Spitzer} and \textit{Herschel}, both sensitivity and confusion noise limitations have impeded our ability to calculate the obscured SFR in typical and low-mass galaxies at $z\sim 1$ and above without relying on stacking many galaxies. MIRI directly traces mid-IR PAH emission up to $z\sim 3$ ($z\sim 2$ with the more sensitive shorter-wavelength bands). Even though converting mid-IR emission to total IR luminosity is model-dependent, its uncertainty is typically lower than that affecting the UV slope methods for moderately and highly obscured galaxies \citep[e.g.][]{shivaei20b}. In this section, we estimate the obscured luminosity fraction from our UV-to-mid-IR SED fits and study its variation as a function of stellar mass and SFR surface density. 

In the previous section, we showed that the PAH fraction reaches a constant average value at metallicities above $\sim 0.5 Z_{\odot}$, and hence using PAH luminosity to estimate total IR luminosity and obscured SFR is robust in this regime. However, as the PAH fraction changes at lower metallicities, instead of a simple PAH-to-IR luminosity conversion, we adopt the full UV-to-mid-IR energy-balanced SED fits to derive the obscured fraction as the ratio of obscured to intrinsic (total) UV luminosity at 1550\,{\AA}: $f_{\rm obsc} = L^{\rm obscured}_{\rm UV}/L^{\rm intrinsic}_{\rm UV} = 1 - (L^{\rm unattenuated}_{\rm UV}/L^{\rm intrinsic}_{\rm UV})$, where the unattenuated UV luminosity is directly constrained by the observed photometry at rest-frame UV and the intrinsic UV luminosity is estimated from the dust-free (intrinsic) best-fit SED model. Our choice of using the UV luminosity to estimate the obscured fraction is because of its multiple implications for high-redshift studies that often only have UV data, but we note that the fractions are in very close agreement with total IR to unattenuated UV luminosity (as was expected). Converting the luminosity fractions to SFR fractions require additional conversions of the luminosities to SFRs, which we decide to avoid in the current paper to mitigate uncertainties.

\subsection{Relationship between obscured fraction and mass}
As most of the metals in the ISM are depleted onto dust grains that attenuate starlight, it is expected that obscured luminosity fraction (ratio of attenuated to intrinsic luminosity) correlates with metallicity, and hence stellar mass \citep[e.g.][]{reddy10, whitaker17}, which is also manifested in the relationship between dust attenuation and mass \citep[e.g.][]{garn10,shivaei20a,bogdanoska20}.

In Figure~\ref{fig:obsc_frac_w17}, the obscured fraction is shown as a function of stellar mass. We investigate the average shape and scatter of the relationship at $M_*> 4\times10^{8}${\Msun}, where the sample is complete. The mass completeness is estimated based on the detection limits of 5.6\,{\um} and 15\,{\um} data and assuming a mass-to-light ratio of 0.6 at 2.2{\um} \citep{mcgaugh14}\footnote{The exact completeness varies with redshift from 0.7 to 2.}.
To study the average trend, we fit the data in three ways: a moving median function\footnote{The moving weighted-average fit shows an identical trend to the moving median curve.}, the functional form of {\fobsc}-$M_*$ suggested by \cite{whitaker17},
\begin{equation} \label{eq:w17}
    f_{\rm obscured} = \frac{1}{1+a~e^{b\log(M_*/M_{\odot})}},
\end{equation}
where $f_{\rm obscured}$ is the obscured fraction, and $a$ and $b$ are the fit parameters, and the following function that best matches the moving median trend:
\begin{equation} \label{eq:mass}
  f_{\rm obsc} = \dfrac{(f_{\rm max}-f_{\rm min})}{2} \times \left({\rm erf}\left(\dfrac{\log(M_*/M_{\odot})-\mu}{\sqrt{2}~\sigma}\right)+1\right) + f_{\rm min}, 
\end{equation}
where erf is the error function, and $f_{\rm max}$, $f_{\rm min}$, $\mu$, and $\sigma$ are the fitting parameters. 
Using an orthogonal distance regression (ODR) fitting procedure ({\tt python scipy.odr}), considering the uncertainties on both the obscured fraction and mass, the fit using Equation~\ref{eq:w17} matches the \cite{whitaker17} curve closely. However, our fit is highly biased by the few highly obscured galaxies with very small uncertainties. The moving (running) weighted-average curve, where the weight is the inverse square of uncertainty in {\fobsc}, also shows how this small population of highly obscured galaxies bias the fit (dotted curve in Figure~\ref{fig:obsc_frac_w17}). The running median (dashed curve) and the Equation~\ref{eq:mass} fit (black line) better represent the behaviour of {\fobsc} as a function of mass.
The best fit parameters are listed in Table~\ref{table:fobsc-mass}.

\begin{table}
\caption{Best-fit parameters of $f_{\rm obscured}-M_*$ for the full sample in Equation~\ref{eq:mass}}             
\label{table:fobsc-mass}     
\centering                  
\begin{tabular}{c c c c} 
\hline\hline  
$f_{\rm max}$& $f_{\rm min}$& $\sigma$& $\mu$\\
\hline 
$0.873\pm 0.015$& $0.273\pm 0.015$& $0.441\pm 0.043$& $9.856\pm 0.026$ \\
\hline 
\end{tabular}
\end{table}

We do not see any significant redshift evolution in this relation, as is shown in the right panel of Figure~\ref{fig:obsc_frac_w17}, in agreement with the findings of \citet{whitaker17} that at a fixed mass the obscured fraction is constant from $z\sim 0.5$ to 2.5.
Based on our best-fit curve (Eq.~\ref{eq:mass}) that matches the moving median, on average, the UV luminosity is dominantly obscured ({\fobsc}$>50\%$) above the stellar mass of $\sim 5\times 10^9$\,\Msun. Interestingly, owing to the large scatter in {\fobsc} at any mass, there are galaxies with dominantly obscured emission even down to $M_*\sim 5\times 10^8$\,\Msun. We investigate the scatter in {\fobsc} at fixed mass in the next section.
Above the stellar mass of $2\times 10^{10}$\,\Msun, more than 80\% of the emission is obscured. However, the SMILES survey is likely not representative of the population of heavily obscured (and rare) galaxies owing to its relatively small area. 

\cite{whitaker17} studied the \textit{Spitzer}/MIPS samples and found that the $f_{\rm obscured}$-$M_*$ relation does not change from $z=0$ to $2.5$. They fit the UV-near-IR photometry with \cite{bc03} models and used the \cite{dh02} template to convert 24{\um} flux densities to total IR luminosities. Then assuming the \cite{kennicutt98} and \cite{bell05} calibrations, they converted UV and IR luminosities to SFRs to estimate $f_{\rm obsc} = \frac{\rm SFR_{IR}}{\rm SFR_{IR}+SFR_{UV}}$. While the two relationships are in relatively good agreement within the scatter of the relation in Figure~\ref{fig:obsc_frac_w17}, there are a few differences between the \cite{whitaker17} study and this work that contribute to some disagreement between the two relationships. \cite{whitaker17} used a single log-averaged IR template from \cite{dh02} (i.e. a fixed IR SED shape) to convert PAH luminosity to total IR luminosity for their sample at $z\sim 0.5-2.5$. While at high metallicities and masses, galaxies may have a more uniform IR SED shape, and hence, the assumption of a single IR SED would be valid, that is not always the case at low metallicities. As we showed in the previous section, below $\sim 0.5 Z_{\odot}$ the PAH fraction decreases uniformly from $z\sim 0-2$. Therefore, using the high-metallicity end calibrations would overestimate the IR luminosities at lower metallicities (and masses; see also e.g. \citealt{shivaei17,shivaei22a}). Therefore, the single IR SED assumption in \citet{whitaker17} might have overestimated their IR luminosity values at lower masses.
Additionally, the different area and sensitivity of SMILES compared to the MIPS/\textit{Spitzer} surveys can play a role in the observed discrepancy between this study and the \citet{whitaker17} study. SMILES is more sensitive to less obscured galaxies at a fixed mass relative to the MIPS samples. In fact, the fit to our data using the \citet{whitaker17} formalism and an ODR fitting procedure (i.e. taking into account the uncertainties on {\fobsc}) closely follows the \citet{whitaker17} curve. This is because a few highly obscured galaxies with very low {\fobsc} uncertainties dominate the fit -- a similar bias that a MIPS-based sample would have. Furthermore, SMILES has a smaller area compared to the \textit{Spitzer}/MIPS surveys, and hence, is likely missing a (rarer) population of massive and highly attenuated galaxies.

\begin{figure*}
        \centering
        \includegraphics[width=\textwidth]{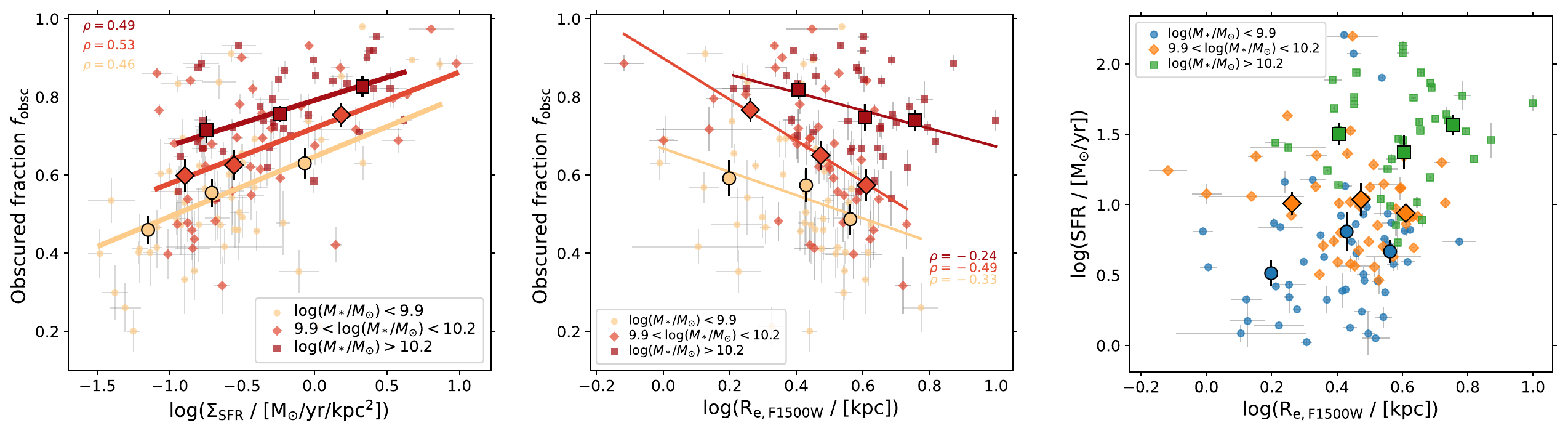}
    \caption{Obscured luminosity fraction as a function of {\Ssfr} (using F1500W sizes) and F1500W effective radius (middle) for the F1500W spatially resolved sample (the filter FWHM is 0.5 arcsec). At a given mass, {\fobsc} increases with {\Ssfr} and decreases with rest-frame mid-IR (dust) radii. Right panel shows SFR versus F1500W size for the same sample. The near constant SFR with ${\rm R_{e,F1500W}}$ in each mass bin indicates that the relationship between {\fobsc} and size is independent of SFR. 
    Individual measurements are shown with small symbols and colour-coded in three stellar mass bins with the same number of galaxies, shown in the legend. In each mass bin, average values in bins of the parameter on the x axis are shown with larger symbols with the corresponding shape and colour of the mass bin. Pearson correlation coefficients ($\rho$) are shown in the corners of the plots with corresponding colours of each mass bin.  Linear fits to the average values of {\fobsc} versus $\log$({\Ssfr}) are shown in the left panel. The fit parameters are in Table~\ref{table:fobsc}. }
    \label{fig:obsc_frac}
\end{figure*}

\subsection{Scatter in {\fobsc}-$M_*$ and relationship with {\Ssfr}}
As was mentioned in the previous section, there is a large scatter in {\fobsc} at a given mass, such that there are galaxies with significant obscured UV emission down to stellar masses of $\sim 5\times 10^8$\,{\Msun}. 
The scatter in the obscured fraction at a given mass increases from 0.1\,dex at $M_*\sim 10^{11}$\,{\Msun} to $\sim 0.5$\,dex at $M_*\sim 10^{9}$\,{\Msun}. 
We calculate secondary correlations in the $f_{\rm obsc}-M_*$ relation with sSFR, age, and SFR surface density using partial correlation coefficients. The surface density is measured using dust sizes from F1500W data (Section~\ref{sec:method-size}). In those calculations the sample is limited to galaxies with spatially resolved discs in F1500W. The spatially resolved sample is limited to galaxies with $M_*>10^{9.5}$\,{\Msun}, and inevitably biased against the extremely compact star-forming galaxies. The stellar mass distribution of the spatially resolved sample above $M_*>10^{9.5}$\,{\Msun} is very similar to that of the parent sample with 50\% of the spatially resolved sample at $M_*<10^{10.2}$\,{\Msun}.
Among the sSFR, age, and SFR surface density parameters, we find a relatively strong correlation between the obscured fraction and SFR surface density at a fixed stellar mass, but no significant correlations with age and SFR at constant mass.

Figure~\ref{fig:obsc_frac} shows the obscured fraction versus SFR surface density ({\Ssfr}, left) and F1500W effective radii (middle). {\Ssfr} is calculated using dust surface area (at rest-frame $8-9$\,{\um}) that closely traces the active star-forming region.
We divide the sample in three stellar mass bins to investigate the trend at a fixed mass. The average values are shown with large symbols in the figure. 

There is a clear correlation between {\fobsc} and {\Ssfr} at all masses with Pearson correlation coefficient of $\sim 0.5$ (p-value $\ll1$). 
We fit the average trends in each mass bin with a linear function. As was expected, the overall normalisation increases with increasing mass; that is, as was shown in the previous figures, {\fobsc} increases with increasing mass.
However, not only does {\fobsc} clearly increase with {\Ssfr} (and decreases with size) at a given mass, but the rate at which {\fobsc} increases with {\Ssfr} (i.e. the slope) appears to be similar across the mass range of the sample. The fit parameters are presented in Table~\ref{table:fobsc}. We discuss the possible explanation of this behaviour in the next Section.
While the overall trend with size (effective radius) is consistent, the slope of the {\fobsc}-size relation varies among the three mass bins and the evolutionary behaviour is not as clear as with {\Ssfr}. We also note that the width of the {\fobsc} distribution decreases as the mass increases, as is shown with the error bars on the average values in Figure~\ref{fig:obsc_frac}. The standard deviation of {\fobsc} in the highest {\Ssfr} bin from low to high mass is 0.20, 0.15, 0.07, respectively.

The right panel of Figure~\ref{fig:obsc_frac} confirms that the {\fobsc} increasing trend with {\Ssfr} is not a byproduct of increasing SFR. 
While there is a general correlation between SFR and size for the full sample, the average SFR does not vary with size in the fixed mass ranges.
Therefore, the underlying cause of {\fobsc} dependence on {\Ssfr} at a given mass is the compactness of the star-forming region, and not only an increase in the rate of star formation.

\begin{table*}
\caption{Fit parameters of the linear fits in Figure~\ref{fig:obsc_frac}. }
\label{table:fobsc}     
\centering                  
\begin{tabular}{c c c c c c} 
\hline\hline   
$N$ & Mass range [$\log(M_*/M_{\odot})$] & \multicolumn{2}{c}{$f_{\rm obscured}$-$\log(\Sigma_{\rm SFR}/M_{\odot}{\rm yr^{-1} kpc^{-2}})$} & \multicolumn{2}{c}{$f_{\rm obscured}$-$\log({\rm R_{e,F1500W}/kpc})$}\\ 
 & & slope & intercept & slope & intercept \\ 
\hline  
50 & $<9.9$  & 0.15$\pm 0.03$ & 0.647$\pm 0.017$ & $-0.29 \pm 0.08$ & $0.668\pm 0.034$\\ 
43 & $9.9-10.2$ & 0.14$\pm 0.02$ & 0.722$\pm 0.011$ & $-0.52\pm 0.08$ & $0.899\pm 0.036$\\
38 & $>10.2$ &0.12$\pm 0.02$ & 0.791$\pm 0.006$& $-0.23\pm 0.04$ & $0.904\pm 0.023$ \\
\hline 
\end{tabular}
\tablefoot{Errors are calculated by taking the median and standard deviation of 200 fits to the randomly perturbed data points within their errors.}
\end{table*}

\begin{figure}
        \centering
        \includegraphics[width=.7\columnwidth]{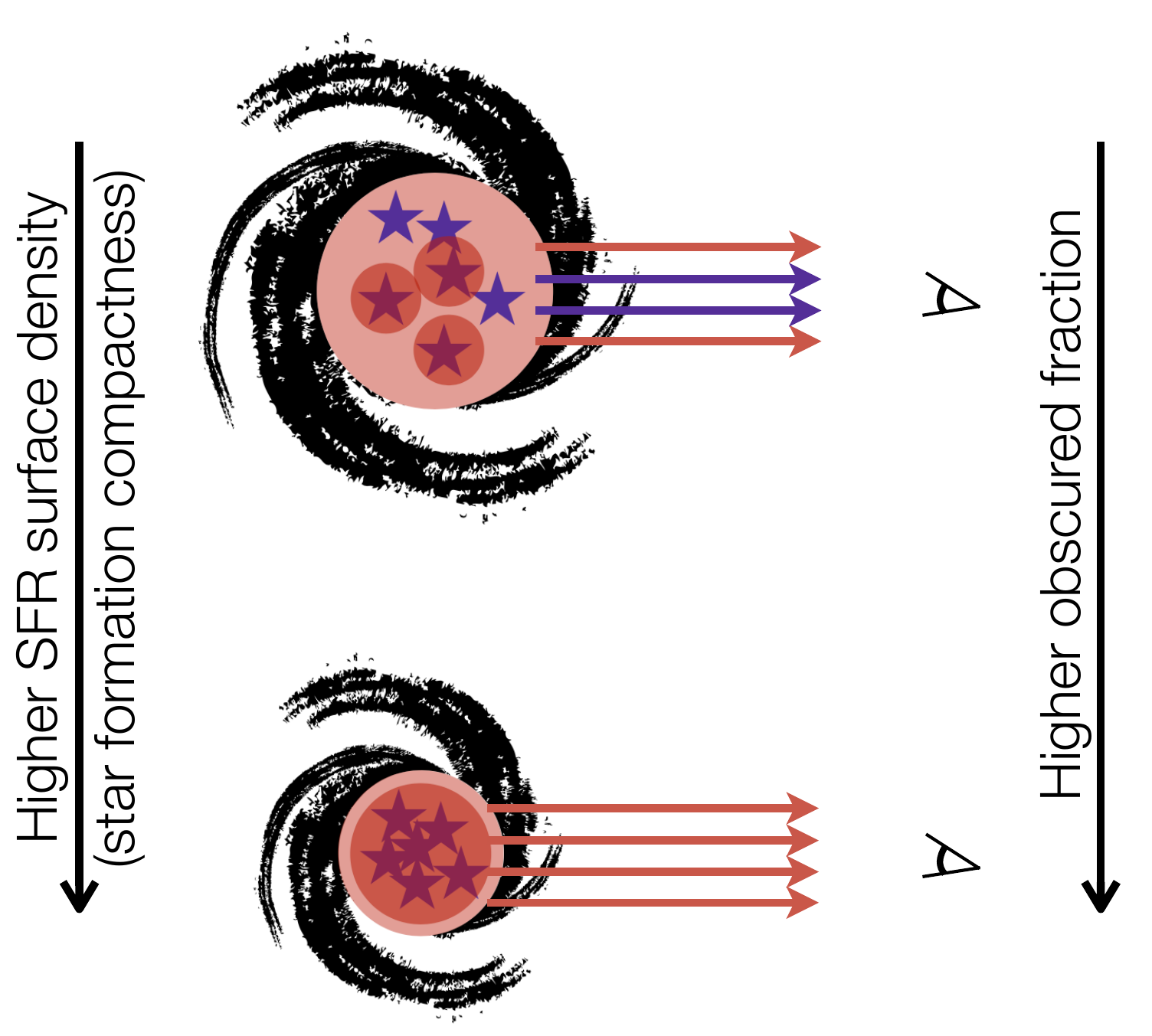} 
    \caption{
    Cartoon illustrating a physical explanation of the correlation between {\fobsc} and star-forming compactness. Stars are shown in blue. Diffuse ISM dust is shown with light red and the `birth cloud' dust with higher column density is shown with a darker red colour. Schematically, less and more dust-obscured light (i.e. obscured by the low and high column density dust, respectively) are shown with blue and red arrows, respectively, reaching to the observer on the right.
    Extended galaxies have a range of dust-star geometries, while compact galaxies contain most of their stellar population in the core with high dust covering fraction, resulting in most of the emitted starlight being obscured by dust.
    }
    \label{fig:fobsc-cartoon}
\end{figure}
\subsection{Morphological evolution and obscured fraction}

Using submm or mm data for samples with generally much higher stellar masses and SFRs than presented here, studies have shown that star formation occurs in more compact regions compared to the stellar mass sizes \citep[e.g.][]{murphy17,jimenez-andrade21,hodge16,elbaz18,fujimoto17,rujopakarn19,hamed23b}. 
Additionally, it has been argued that starburst galaxies on the main-sequence with particularly short depletion timescales exhibit higher SFR surface densities with more compact mm dust emission \citep{elbaz18,gomez-guijarro22} and radio emission \citep{jimenez-andrade19}, and hotter dust \citep{gomez-guijarro22} compared to other main-sequence galaxies at the same mass and redshift.
In the regime of lower SFR surface density ($\log(\Sigma_{\rm SFR}/M_{\odot}{\rm yr^{-1}}~{\rm kpc^{-2}})\sim -1$ to 1 ) and mass ($\log(M_*/M_{\odot})\sim 9.5$ to 11), we observe the same relation for purely star-forming galaxies (no AGN contamination) that PAH dust sizes are more compact with higher SFR surface densities at a fixed mass (right panel of Figure~\ref{fig:obsc_frac}). 

We also see that the obscured emission is more dominant in compact galaxies with higher {\Ssfr}. This observed trend is illustrated by a simple cartoon in Figure~\ref{fig:fobsc-cartoon}. For an extended star-forming galaxy the dust-star geometry is more complex: some population of stars are only affected by the diffuse ISM dust, while others can be trapped in their birthcloud dust and highly attenuated. If the star-forming region is compressed into a small area, the galaxy will have a more homogeneous dust distribution with higher covering fraction that attenuates most of the light emitted from the stars. Hence, the obscured fraction of the emitted light is higher in galaxies with compact star-forming regions. This explanation works for closed-box systems. However, strong outflows and feedback may clear pathways through the ISM of the compact star-forming regions and expose the starlight without attenuation. This may explain the larger variation in {\fobsc} in the low mass bins at any {\Ssfr}: low-mass galaxies have highly stochastic star formation histories \citep{hopkins14}, and massive stars produced in the vigorous starbursts can destroy their birth clouds via feedback (e.g. winds, supernovae; \citealt{naidu22}) to make sightlines clear of dust. This will generate a diversity of {\fobsc} from galaxy to galaxy at low masses, while in massive galaxies dust stays in the compact core with high covering fraction. In our data, the scatter (standard deviation) of {\fobsc} in the lowest mass bin is $15-20$\% at a fixed {\Ssfr} and decreases to $8-12$\% in the highest mass bin.

\begin{figure*}
        \centering
        \includegraphics[width=\textwidth]{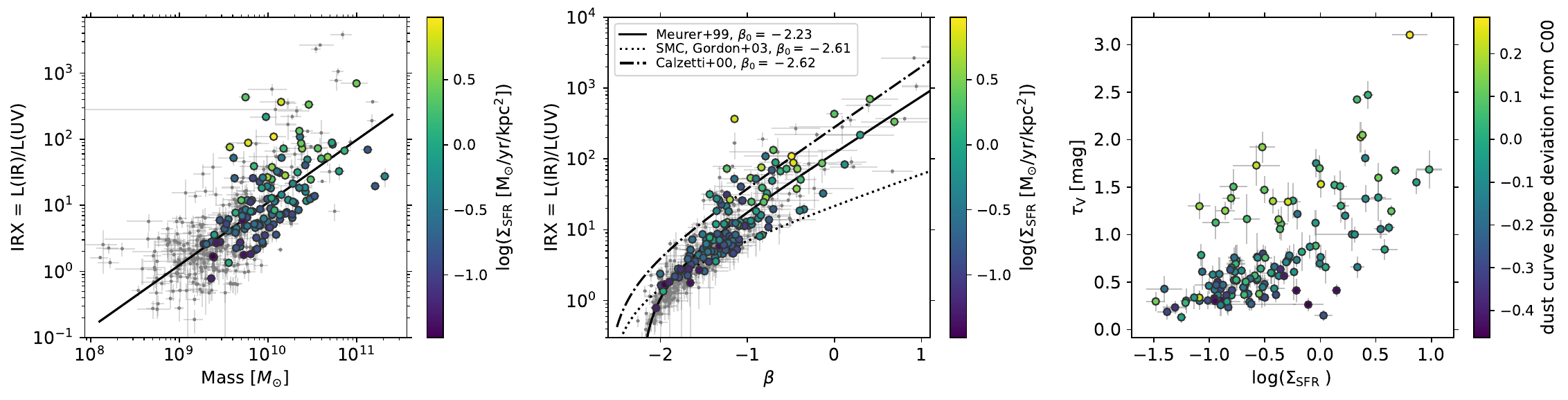} 
    \caption{
IRX (ratio of IR to UV 1550\,{\AA} luminosity) as a function of stellar mass (left panel) and UV stellar continuum slope (middle panel). Galaxies with spatially resolved morphology in F1500W are colour-coded by their {\Ssfr}. A linear fit to $\log({\rm IRX})-\log(M_*/M_{\odot})$ is shown in the left panel (slope of $0.96\pm 0.06$ and intercept of $-8.47\pm 0.61$). In the middle panel, we show three IRX-$\beta$ relations in the literature adopted from \cite{reddy18a} assuming a constant star formation history and an age of 100\,Myr for stellar populations: the original \citet{meurer99} relation, the starburst curve of \citet{calzetti00} and the SMC curve of \citet{gordon03} assuming 0.14 Solar metallicity. A higher metallicity system will result in a redder intrinsic UV slope ($\beta_0$), shifting the curves to the right of the diagram (e.g. the Calzetti curve for a 1.4\,$Z_{\odot}$ population and $\beta_0 =-2.4$ resembles the \citealt{meurer99} curve). Galaxies with higher {\Ssfr} at a given mass or $\beta$ have higher IRX and a shallower attenuation curve. Right panel: $V$-band dust optical depth versus {\Ssfr}, colour-coded with the attenuation curve slope defined as the deviation from the \cite{calzetti00} (the formalism of \citealt{kriek13}), where zero is the Calzetti curve slope. The two parameters are highly correlated with $\rho = 0.65$ (p-value $\ll1$). The  shallower attenuation curve of galaxies with high IRX at a given mass or $\beta$ can be explained by the radiative transfer effects that are a result of their high {\Ssfr}: higher optical depths and lower fractions of unobscured evolved stars. 
    }
    \label{fig:irx}
\end{figure*}

The relationship of obscured fraction with compactness of star-forming region can also explain the spread in the locus of galaxies in the IRX-$\beta$ diagram. The IRX-$\beta$ relation is an empirical calibration of UV stellar continuum slope ($\beta$) based on the ratio of IR to unobscured UV luminosity (IRX; \citealt{meurer99}). The location of galaxies in this diagram depends on the slope of the attenuation curve in the UV and the intrinsic UV slope \citep[e.g.][]{reddy18a,shivaei20b,salim20,hamed23}. Figure~\ref{fig:irx} shows the distribution of our sample in the IRX-mass and IRX-$\beta$ diagrams. Galaxies with higher {\Ssfr} at a given mass or $\beta$ have higher IRX, and preferentially shallower attenuation curves similar to that of the \cite{calzetti00} curve (middle panel; similar trends are also seen at $z\sim 0$, \citealt{johnson07}). The preference for shallower curves is potentially due to both a higher average optical depth and a lower fraction of unobscured evolved stars. Radiative transfer models have shown that a high dust optical depth flattens the dust attenuation curve \citep{chevallard13} and the right panel of Figure~\ref{fig:irx} shows that {\Ssfr} in our sample is correlated with the dust optical depth (Pearson $\rho = 0.65$ with p-value $\ll1$). Additionally, as is shown in the schematic Figure~\ref{fig:fobsc-cartoon}, galaxies with more concentrated star-forming regions (higher {\Ssfr}) have lower fraction of evolved stars unobscured by dust, as these stars would not have sufficient time to migrate away from their dusty birth clouds. This dust-star geometrical effect also tends to flatten the dust attenuation curve of galaxies in which the evolved stars dominate the optical luminosity \citep{narayanan18}. The inclination (axis ratio) of galaxies can also play a role in the observed trends, which will be investigated in future studies.

\section{Conclusions}\label{sec:summary}

In this study, we have used SMILES, the largest 5-25\,{\um} MIRI survey in the first two observing cycles of JWST, to study the dust-obscured and PAH grain properties of 443 galaxies at $z=0.7-2.0$ (median redshift of $z\sim 1.3$){, in which 216 of them have well-constrained {\qpah} values, dubbed the `robust {\qpah}' sample}. The exceptional sensitivity, angular resolution, and mid-IR wavelength coverage offered by SMILES enables unambiguous detection of dust emission from star-forming regions in galaxies, reaching mass and SFR an order of magnitude lower than was achievable with JWST's predecessor, \textit{Spitzer}. Augmented by the deep NIRCam (JADES) and HST (CANDELS) data in GOODS-S, we performed SED fitting on the extremely well-covered photometry spanning from rest-frame UV to mid-IR. We have studied the PAH to dust mass fractions ({\qpah}) and dust-obscured fraction of UV luminosity ({\fobsc}) in relation to galaxy parameters such as stellar mass, metallicity, and SFR surface density. The sample contains main-sequence galaxies spanning a range from $\sim 10^{8.4} - 10^{11.3}$\,{\Msun}. Our key findings are summarised below.

\begin{itemize}
    \item We find a strong correlation between the mass fraction of dust in PAHs (i.e. PAH mass fraction or {\qpah}) and stellar mass. A similar correlation is seen between the PAH luminosity fraction relative to the total IR luminosity and mass. This correlation positions the {\qpah} value of the majority of galaxies at $z\sim 1.3$ between the median values of low- and high-metallicity galaxies at $z\sim 0$ \citep{draine07b}.

    \item Using the FMR, we estimated the gas metallicity of the sample and compared the {\qpah}-metallicity relation at $z\sim 1.3$ with that at $z\sim 0$. Despite the uncertainties in the $z\sim 0$ metallicities due to strong line calibrations ($\sim 0.2$\,dex, \citealt{aniano20} compared to the $\sim 0.05$\,dex metallicity scatter from the FMR), we observe a strong agreement in the {\qpah}-metallicity trends from $z\sim 0$ to 2. In this relationship, {\qpah} remains relatively constant at $\sim 3.4$\% above $\sim 0.5\,Z_{\odot}$, confirming the reliability of PAH luminosity as a robust calibrator for estimating the total IR luminosity (and similarly, the PAH mass to dust mass) in metal-rich galaxies. Below $0.5\,Z_{\odot}$, the PAH fraction declines sharply, reaching a very low value of $< 1$\% at metallicities below $0.3\,Z_{\odot}$. This redshift-invariant nature of the {\qpah}-metallicity relation below $\sim 0.5\,Z_{\odot}$ could potentially serve as a basis for calibrating the PAH luminosity fraction as a metallicity indicator in this range. This would be an efficient way of estimating metallicities from relatively cost-effective imaging surveys, with MIRI applicable up to $z\sim 2$ and potential future far-IR probes such as SPICE (\citealt{urry23}, Bonato et~al., in prep) or PRIMA \citep{moullet23} extending capabilities to $z\sim 7$.

    \item We observe a strong correlation between the fraction of obscured UV emission (the ratio of obscured to total UV luminosity) and stellar mass that does not change with redshift. Galaxies with stellar masses above $10^{9.6}$\,{\Msun} have, on average, more than half of their light obscured by dust. In the stellar mass range of $\sim 10^{8.5}$ to $10^{11}$\,{\Msun}, the obscured fraction increases from $\sim 15\%$ to more than $90\%$. 

    \item We determined dust sizes using F1500W images, which capture the $\sim 5-8.8$\,{\um} dust and PAH emission within the redshift range of $z=0.7-2$. The spatial resolution of MIRI allows us to obtain sizes of galaxies at stellar masses $\sim 10^{9-11}$\,{\Msun} (the median mass within the spatially resolved sample is $10^{10}$\,{\Msun}). Our findings indicate that, at a given mass, galaxies with compact dust regions (indicative of active star formation) exhibit higher SFR surface densities ({\Ssfr}). This suggests shorter depletion timescales in compact galaxies. Additionally, at the same mass, galaxies with higher {\Ssfr} also exhibit higher obscured luminosity fractions. However, among lower-mass galaxies there is a broader range of {\fobsc} at a fixed {\Ssfr}.

    \item We explain the observed correlation between {\fobsc} and {\Ssfr} at a constant mass by attributing it to the increase in the dust covering fraction as the star-forming region becomes more compact. 
    Simultaneously, the observed increase in the scatter of {\fobsc} at low masses compared to that in more massive bins can be attributed to the more bursty nature of star formation in lower-mass galaxies. Feedback mechanisms may contribute to the wider range of dust covering fractions at low masses for a given {\Ssfr}.

    \item Similarly, we find that for a given mass or UV continuum slope ($\beta$), galaxies with higher attenuation quantified by the IRX (IR to unobscured UV luminosity) parameter tend to possess higher {\Ssfr}, positioning them within the region of the IRX-$\beta$ diagram that aligns with shallower attenuation curves. Such shallow attenuation curves may arise from both elevated average dust optical depths and a compact dust-star geometry (with reduced fraction of unobscured evolved stars) at high SFR surface densities.
    
\end{itemize}

We are just at the beginning of an exciting journey with MIRI to unveil the properties of dust beyond the local Universe, in an unexplored SFR-mass regime. In the future, larger samples could help to solidify the trends presented here. Additionally, more sophisticated spatial analysis of the mid-IR data, as unbiased probes of intense star formation, will be invaluable in shedding light on the morphological evolution of galaxies.

\begin{acknowledgements}
IS thanks the members of the JWST/MIRI instrument team for their exceptional efforts and for providing an outstanding experience  during the commissioning period of JWST, which fostered numerous fruitful discussions and significantly enhanced the quality of data reduction in this study. IS also thanks Karin Sandstrom and Joel Leja for their insightful discussions during the scientific development of this work. Additionally, IS acknowledges the contribution of Andras G{\'a}spar to the construction of the F560W PSF utilised in this research.

This work was supported in part by NASA grant NNX13AD82G. Part of this research has been funded by Atracc{\' i}on de Talento Grant No.2022-T1/TIC-20472 of the Comunidad de Madrid, Spain. AJB and AC acknowledges funding from the `FirstGalaxies' Advanced Grant from the European Research Council (ERC) under the European Union’s Horizon 2020 research and innovation program (Grant agreement No. 789056). The work of CCW is supported by NOIRLab, which is managed by the Association of Universities for Research in Astronomy (AURA) under a cooperative agreement with the National Science Foundation. PGP-G acknowledges support from grant PID2022-139567NB-I00 funded by Spanish Ministerio de Ciencia e Innovaci\'on CIN/AEI/10.13039/501100011033, FEDER {\it Una manera de hacer Europa}. SA acknowledges support from the JWST Mid-Infrared Instrument (MIRI) Science Team Lead, grant 80NSSC18K0555, from NASA Goddard Space Flight Center to the University of Arizona. 

This work is based on observations made with the NASA/ESA/CSA \textit{James Webb} Space Telescope. The data were obtained from the Mikulski Archive for Space Telescopes at the Space Telescope Science Institute, which is operated by the Association of Universities for Research in Astronomy, Inc., under NASA contract NAS 5-03127 for JWST.  These observations are associated with program PID 1207, 1080, 1081, 1895, 1220, 1286, 1287, 1963.
Based on observations made with the NASA/ESA \textit{Hubble} Space Telescope, and obtained from the Hubble Legacy Archive, which is a collaboration between the Space Telescope Science Institute (STScI/NASA), the Space Telescope European Coordinating Facility (ST-ECF/ESAC/ESA) and the Canadian Astronomy Data Centre (CADC/NRC/CSA).
\end{acknowledgements}

\bibliographystyle{aa}
\bibliography{bibliography} 

\end{document}